\RequirePackage{fix-cm}
\documentclass[smallextended]{svjour3}       
\smartqed  
\usepackage{graphicx}
\usepackage[table,xcdraw,dvipsnames]{xcolor}
\usepackage{hyperref} 
\usepackage{xurl}
\usepackage{lscape}
\usepackage{multirow}
\usepackage[utf8]{inputenc}
\usepackage{makecell}
\usepackage{tcolorbox}
\usepackage{natbib}
\usepackage{makecell}
\usepackage{diagbox}
\usepackage{adjustbox}
\usepackage{tabularx}
\usepackage{subfigure}
\definecolor{diffstart}{HTML}{004db5}
\definecolor{diffincl}{HTML}{00990c}
\definecolor{diffrem}{HTML}{ad0000}
\definecolor{codegreen}{rgb}{0,0.6,0}
\definecolor{codegray}{rgb}{0.5,0.5,0.5}
\definecolor{codepurple}{rgb}{0.58,0,0.82}
\definecolor{backcolour}{rgb}{0.95,0.95,0.92}
\usepackage{listings}
  \lstdefinelanguage{diff}{
	basicstyle=\ttfamily\extrabold\scriptsize,
	morecomment=[f][\color{diffstart}]{@},
	morecomment=[f][\color{diffincl}]{+},
	morecomment=[f][\color{diffrem}]{-},
        keepspaces=true,
	identifierstyle=\color{black},
    numbers=left,    
  }
\lstdefinestyle{mystyle}{
    backgroundcolor=\color{backcolour}, frame=lrtb,
    commentstyle=\color{codegreen},
    keywordstyle=\color{magenta},
    numberstyle=\tiny\color{codegray},
    stringstyle=\color{codepurple},
    basicstyle=\ttfamily\footnotesize,
    breakatwhitespace=false,         
    breaklines=true,                 
    captionpos=b,                    
    keepspaces=true,                 
    numbers=left,                    
    numbersep=5pt,                  
    showspaces=false,                
    showstringspaces=false,
    showtabs=false,                  
    tabsize=2,
}
\usepackage{algorithmic}
\usepackage{textcomp}
\usepackage{xcolor}
\usepackage{url}
\usepackage{multirow}
\usepackage{balance}
\usepackage{stfloats}
\usepackage{booktabs}
\usepackage{makecell}
\usepackage{array}
\usepackage{tabularx}
\usepackage{booktabs}
\usepackage[normalem]{ulem}
\usepackage{enumerate}
\usepackage[normalem]{ulem}
\usepackage[utf8]{inputenc}
\usepackage{amsmath}  
\usepackage{comment}
\usepackage{enumitem}
\usepackage{tcolorbox}
\usepackage{balance}
\usepackage{mathtools}
\usepackage{xspace}
\usepackage{eucal}
\usepackage{amsfonts}
\usepackage{color}
\usepackage{bbding}
\usepackage{graphicx}
\usepackage{hyperref}
\usepackage{colortbl}
\usepackage{lscape}
\usepackage{adjustbox}

\usepackage{framed}
\usepackage{arydshln}
\definecolor{dark-red}{RGB}{255,0,0}
\definecolor{dark-green}{RGB}{0,200,0}
\definecolor{intnull}{RGB}{213,229,255}
\definecolor{rintnull}{RGB}{255,229,213}
\definecolor{yintnull}{RGB}{255,213,229}
\definecolor{xintnull}{RGB}{255,220,120}
\definecolor{zintnull}{RGB}{240,200,180}

\newcommand{\revised}[1]{\textcolor{black}{#1}}

\newcommand{\find}[1]{
\begin{tcolorbox}[size=fbox,boxsep=2mm,boxrule=0pt,top=0pt,bottom=0pt,colback=blue!5!white]
\em #1
\end{tcolorbox}
}

\AtBeginDocument{%
  \providecommand\BibTeX{{%
    \normalfont B\kern-0.5em{\scshape i\kern-0.25em b}\kern-0.8em\TeX}}}

\begin{document}

\title{Bridging Expert Knowledge with Deep Learning Techniques for Just-In-Time Defect Prediction}

\titlerunning{Bridging Expert Knowledge with Deep Learning}        

\author{Xin Zhou \and
 DongGyun Han \and
David Lo}


\institute{Xin Zhou, David Lo \at
School of Computing and Information Systems, Singapore Management University \\ 80 Stamford Rd, Singapore \\ \email{\{xinzhou.2020,davidlo\}@smu.edu.sg}
\and
DongGyun Han \at
Department of Computer Science, Royal Holloway, University of London \\ Egham, Surrey TW20 0EX, United Kingdom (UK) \\ \email{donggyun.han@rhul.ac.uk}    %
}
\date{Received: date / Accepted: date}

\maketitle
\begin{abstract} 
Just-In-Time (JIT) defect prediction aims to automatically predict whether a commit is defective or not, and has been widely studied in recent years. In general, most studies can be classified into two categories: 
1) simple models using traditional machine learning classifiers with hand-crafted features, 
and 
2) complex models using deep learning techniques to automatically extract features from commit contents.
Hand-crafted features used by simple models are based on expert knowledge but may not fully represent the semantic meaning of the commits.
On the other hand, deep learning-based features used by complex models represent the semantic meaning of commits but may not reflect useful expert knowledge.
Simple models and complex models seem complementary to each other to some extent.
To utilize the advantages of both simple and complex models, we propose a model fusion framework that adopts both early fusions on the feature level and late fusions on the decision level.
We propose SimCom++ by adopting the best early and late fusion strategies.
The experimental results show that SimCom++ can significantly outperform the baselines by 5.7--26.9\%.
In addition, our experimental results confirm that the simple model and complex model are complementary to each other.

\end{abstract}

\section{Introduction}
\label{sec:intro}


Modern software development tends to release software in a short period~\citep{Mockus2000PredictingRO, Nan2009ImpactOB}, although developers often have limited testing resources~\citep{Mockus2000PredictingRO, Kamei2015StudyingJD}. \revised{For example, the availability of skilled testers can be limited.}
Such limited testing resources and tight development schedules often lead to introducing software defects~\citep{Cabral2019ClassIE}.
To address the challenge, many defect prediction approaches have been proposed to help developers to narrow down the testing and debugging scope to software components that are highly likely to contain defects~\citep{DAmbros2011EvaluatingDP,Kamei2016DefectPA}.
The Just-in-Time (JIT) defect prediction approach is a specific type of defect prediction approach that can identify the possible defective code changes as soon as they are submitted, which can avoid the defects being merged to the codebase. 
Besides, JIT defect prediction approaches can give fine-grained predictions at the change level, while traditional defect prediction approaches only give coarse predictions at the file level or module level~\citep{DAmbros2011EvaluatingDP,Turhan2008OnTR}.
Moreover, when a commit is submitted for a review, developers need to comprehend it to see whether to integrate it into the code base or not. JIT defect prediction approaches can help to flag cases that require developers’ more scrutiny and thus guide the comprehension process to focus on the more risky cases.

In the past decades, JIT defect prediction has attracted many researchers and many approaches have been proposed~\citep{Mockus2000PredictingRO,Kamei2013ALE,Yang2015DeepLF,Liu2017CodeCA,Yang2017TLELAT,Chen2018MULTIME,Young2018ARS,Cabral2019ClassIE,Hoang2019DeepJITAE,Yan2020JustInTimeDI,Hoang2020CC2VecDR,Zeng2021DeepJD,zhou2022simple}.
\revised{In general, most work can be classified into two categories: 
1)  \textbf{Simple models:} The models involve the use of traditional machine learning (ML) classifiers with manually designed features.
2) \textbf{Complex models:} The models employ deep learning (DL) techniques to automatically extract features from various elements within commits, such as commit logs and code changes. Here, we refer to traditional machine learning as ML and deep learning as DL for simplicity.
ML JIT models are considered simple because they have a limited number of parameters to train and are very fast.
ML JIT models are considered simple because they have a limited number of parameters to train and are fast. In contrast, DL JIT models are considered complex due to their large number of parameters, which demand more time for training and evaluation than ML JIT models.
The majority of existing work are simple models, which rely heavily on hand-crafted change-level features~\citep{Mockus2000PredictingRO,Kamei2013ALE,Yang2015DeepLF,Liu2017CodeCA,Yang2017TLELAT,Chen2018MULTIME,Young2018ARS,Cabral2019ClassIE,Yan2020JustInTimeDI,Zeng2021DeepJD}.}

There exists a saga of simple vs. complex in the JIT defect prediction task (and beyond)~\citep{Hoang2019DeepJITAE,Zeng2021DeepJD, Fu2017EasyOH, Majumder2018500TF}.
Even though complex models are slower than simple models, complex models usually show their superiority over simple models in predictive power. For instance, DeepJIT (a complex model) outperforms prior simple models by 7--9\% in terms of Area Under the Receiver Operating Characteristic Curve (AUC-ROC)~\citep{Hoang2019DeepJITAE}. 
However, recently, Zeng et al. proposed a simple model namely LApredict that achieved  2.6\% improvements over complex models in AUC-ROC~\citep{Zeng2021DeepJD}.

Simple models use a set of hand-crafted features to represent a commit. These features are based on expert knowledge and describe some properties of commits such as code change size, code change diffusion, and history of code change.
However, those hand-crafted features may not fully represent the semantic meaning of the actual code changes in commits.
For instance, two commits with different purposes and defectiveness labels may still have the same hand-crafted features (e.g. code change size)~\citep{Hoang2019DeepJITAE}.

On the other hand, complex models use features that are automatically extracted from the raw contents of commits via DL techniques.
These features represent the semantic meaning of commits~\citep{Hoang2019DeepJITAE}, which are proven useful in various software engineering tasks including JIT defect prediction~\citep{Hoang2019DeepJITAE,Wang2016AutomaticallyLS, Nguyen2015GraphBasedSL, Hindle2016OnTN}.
\revised{However, complex models (deep learning-based models) are well-known for their “black-box” nature. The “black-box” nature prevents us from knowing the meanings of the features automatically extracted by complex models.}
In addition, these automatically extracted features may not leverage useful expert knowledge. 
For instance, code files changed by many developers in the past may include defects~\citep{Matsumoto2010AnAO}.
However, the content in a single commit does not explicitly contain information about the number of developers who have changed the modified files of this commit in the past.
\revised{Although the commit history (i.e., all the commits of a software) contain this information, the commit history is too long for deep learning models to deal with. This is because deep learning models often demand substantial GPU resources, and longer inputs require even more extensive GPU resources.
}

In terms of features, the simple model and complex model seem complementary to each other, which motivates us to combine these two models to build a better JIT defect prediction model.
However, these two features have different properties:
hand-crafted features are a list of numbers and each number has a clear meaning while DL features are vectors whose elements are not mutually exclusive and do not have clear interpretations~\citep{Goodfellow2015DeepL}.
Besides, the commonly used classifiers for hand-crafted features (e.g. Logistic Regression) and DL features (e.g. the fully connected layers) are also different.

To better make use of two kinds of features, in this paper, we propose a framework that can effectively combine the advantages of the two features and build an approach namely SimCom++ by adopting the framework.
SimCom++ is based on our proposed simple and complex models that are inspired by previous work~\citep{Yang2017TLELAT, Hoang2019DeepJITAE, Zeng2021DeepJD}.
The simple and complex models are combined on both the prediction level (i.e. fusing the prediction scores from different models) and the feature level (inserting hand-crafted features into DL features).
Based on the framework, this paper investigates the following research questions:

\begin{itemize}
    \item RQ1: How effective are expert knowledge-based and deep learning-based JIT defect prediction techniques \revised{prior to fusion}?

    \item RQ2: How different are the predictions of expert knowledge-based and deep learning-based approaches?

    \item RQ3: What are the best model fusion strategies in the JIT defect prediction?

     \item \revised{RQ4: Why does the model fusion lead to better effectiveness?}

\end{itemize}

Answering these questions helps fill the gaps of why and how to combine expert knowledge with deep learning approaches. 
First, by checking the effectiveness of expert knowledge-based and deep learning-based tools, we can identify the most competitive models and regard those models as candidate models.
Second, it is important to consider how different the predictions produced by candidate models are. Different predictions may indicate that models have learned different aspects and properties and in this case, combining them may lead to improvements.
Third, given the various ways to combine models, it is important to identify the best combination strategies and report the model performance of the combined model.
Lastly, we aim to give several observations to explain how our framework works.
To address these questions, we present a new model fusion framework that can make advantage of both expert knowledge and deep learning models and achieve new state-of-the-art performance. 

The experimental results show that our approach SimCom++ can significantly outperform baselines by 5.7\%, 12.5\%, and 17.9\% in terms of AUC-ROC, AUC-PR, and F1-score respectively.
Besides, SimCom++ shows better performance in a diverse set of projects than the simple model only or the complex model only.

\vspace{0.3cm}
Our main contributions can be summarized as follows:
\begin{itemize}
    \item{
We propose a comprehensive framework to combine a simple model (expert knowledge) and a complex model (deep learning features). 
}
    \item{
We conduct a study to evaluate our proposed approach on a large and diverse dataset with a total of 94,818 commits. The results show that our approach outperforms the baselines by a large margin. 
}   
    \item{\revised{We conducted an extensive examination to identify the effective combined model that harnesses the distinct information captured by simple and complex models, underscoring the pivotal role of this integration in enhancing predictive capabilities in defectiveness.}}
\end{itemize}

The rest of the paper is organized as follows. 
Section 2 provides background information on existing JIT defect prediction approaches. 
Section 3 introduces our proposed framework to combine expert knowledge and deep learning models. 
Section 4 describes our experimental settings and research questions. 
Section 5 reports our experimental results to answer the research questions.
Section 6 discusses our findings.
Section 7 presents the related work. 
Section 8 concludes the paper with future work.
\section{Background}

The Just-in-Time (JIT) defect prediction approach is to predict if a commit is defective or not. 
We briefly introduce the well-known simple (expert knowledge-based) and complex (deep learning-based) JIT defect prediction models.

\vspace{-0.3cm}
\subsection{Expert Knowledge-based Models}

As presented in Table 1, simple models rely heavily on hand-crafted change-level features where expert knowledge about the defectiveness of commits is embedded in the hand-made features proposed in the literature~\citep{Mockus2000PredictingRO,Kamei2013ALE,Yang2015DeepLF,Liu2017CodeCA,Yang2017TLELAT,Chen2018MULTIME,Young2018ARS,Cabral2019ClassIE,Yan2020JustInTimeDI,Zeng2021DeepJD}.
In general, researchers usually feed the hand-crafted features to a machine learning classifier (e.g. Logistic Regression and Decision Tree) to predict if commits are defective.
We introduce several popular expert knowledge-based JIT defect prediction approaches in this subsection.

\vspace{0.1cm}
\noindent\textbf{LR-JIT}~\citep{Kamei2013ALE} is a classic simple model that is proposed by Kamei et al.~\citep{Kamei2013ALE}, which integrates 14 commit-level hand-crafted features with the logistic regression model to predict if a commit is defective or not.
These 14 features describe the properties of code changes, such as the size, diffusion, history, and author experience of code changes.
LR-JIT has been widely adopted as an evaluation baseline for many previous JIT defect prediction studies~\citep{Yang2015DeepLF, Yang2016EffortawareJD, Zeng2021DeepJD}.

\vspace{0.1cm}
\noindent\textbf{DBN-JIT}~\citep{Yang2015DeepLF} is proposed by Yang et al.~\citep{Yang2015DeepLF}  which uses Deep Belief Network (DBN) to extract high-level information from the commit-level defect prediction features mentioned above. Although DBN-JIT adopted Deep Learning techniques (i.e. Deep Belief Network), its goal is still to make better use of hand-crafted features and not extract features automatically from the contents of commits. Thus, we categorize it as a simple model (i.e. relying heavily on hand-crafted features) in the context of this paper.

\vspace{0.1cm}
\noindent\textbf{LApredict}~\citep{Zeng2021DeepJD} is a simple model that only makes use of \textit{the number of added lines} feature with the logistic regression classifier.
In other words, LApredict gives defectiveness prediction only based on the number of code lines added in a commit: the more lines of code are added in a commit, the higher is its probability of being defective.
LAPredict~\citep{Zeng2021DeepJD} has outperformed the other existing JIT defect prediction approaches regardless of its simplicity.

\vspace{0.1cm}

\revised{\noindent\textbf{XGBoost}~\citep{Zeng2021DeepJD} is an ensemble of traditional ML models (Decision Trees). It builds an ensemble of decision trees sequentially, where each new tree corrects the errors made by the previous ones. This iterative process continues until a stopping criterion is met.
Because XGBoost still relies on hand-crafted features and does not use any Deep Learning models to automatically extract features from commit contents, we still regard XGBoost as the simple model.
In this paper,  we use XGBoost as a baseline to show the superiority of combining both expert knowledge-based and deep learning models (e.g. our SimCom++) over simply ensembling multiple traditional ML models (e.g. XGBoost).
As we are the first to use XGBoost in the JIT defect prediction, we tuned the hyperparameters of XGBoost
to make the best use of it in this task. Specifically, we tuned the following important hyperparameters of XGBoost based on the performance of the validation sets:}
\begin{itemize}
    \item \revised{Minimum loss reduction: It controls the model whether to make a further partition on a leaf node of the tree. The larger the value is, the more conservative the algorithm will be. The default value is 0.0, We did a grid search on the potential values \textit{min-loss-reduction=[0.0, 0.1, 0.2, 0.3]}.}
    \item \revised{Maximum depth of a tree: It indicates the minimum number of samples required to be at a leaf node. The default value is 6. We did a grid search on the potential values \textit{max-depth=[3, 6, 9, 12]}.}
    \item \revised{Minimum child weight: If the tree partition step results in a leaf node with the sum of instance weight less than the minimum child weight, then the building process will give up further partitioning. The default value is 1. We did a grid search on the potential values \textit{min-child-weight=[0.5, 1, 1.5, 2]}.}
\end{itemize}
\revised{We iterated to explore all hyperparameter combinations to identify a set of hyperparameters that achieves the best performance on the validation sets.  After the experiments, we found the best hyperparameters of XGBoost are: \textit{min-loss-reduction=0.0, max-depth=3, and min-child-weight=1.5}. and use the best hyperparameters in training XGBoost.}

\begin{table}[t]
\caption{Widely used 14 hand-crafted features for code changes}
	\centering
	\scriptsize
    \begin{tabularx}{\linewidth}{lX}
    \toprule
    \textbf{Name} & \textbf{Description}                                          \\
    \midrule
    NS            & The number of modified subsystems~\citep{Mockus2000PredictingRO}                               \\
    ND            & The number of modified   directories~\citep{Mockus2000PredictingRO}                            \\
    NF            & The number of modified files~\citep{Nagappan2006MiningMT}                                   \\
    Entropy       & Distribution of modified code across each file~\citep{DAmbros2010AnEC, Hassan2009PredictingFU}                \\
    LA            & Lines of code added~\citep{Moser2008ACA, Nagappan2005UseOR}                                             \\
    LD            & Lines of code deleted~\citep{Moser2008ACA, Nagappan2005UseOR}                                           \\
    LT            & Lines of code in a file before the change~\citep{Koru2009AnII}                     \\
    FIX           & Whether or not the change is a defect fix~\citep{Guo2010CharacterizingAP, Purushothaman2005TowardUT}                     \\
    NDEV          & The number of developers that changed the modified files~\citep{Matsumoto2010AnAO}      \\
    AGE           & The average time interval between the last and current change~\citep{Graves2000PredictingFI} \\
    NUC           & The number of unique changes to the modified files~\citep{DAmbros2010AnEC, Hassan2009PredictingFU}            \\
    EXP           & Developer experience~\citep{Mockus2000PredictingRO}                                            \\
    REXP          & Recent developer experience~\citep{Mockus2000PredictingRO}                                     \\
    SEXP          & Developer experience on a subsystem~\citep{Mockus2000PredictingRO}                             \\
    \bottomrule
    \end{tabularx}
    \label{hand_feature}
\end{table}

\subsection{Deep Learning-based Models}

Dissimilar to simple models that depend on hand-crafted features derived from expert knowledge, another stream of works focuses on automatically extracting features from the content of commits by leveraging Deep Learning techniques. Recently, two complex models have been proposed, namely DeepJIT~\citep{Hoang2019DeepJITAE} and CC2Vec~\citep{Hoang2020CC2VecDR}. We briefly introduce those two Deep Learning-based JIT defect prediction approaches in this subsection. In addition, though not designed for this task, we also include a popular pre-trained model of code (i.e., CodeBERT~\citep{codebert}) because it has shown its great effectiveness in a wide range of downstream tasks such as code search and code summarization~\citep{codebert,zhou2021assessing,lu2021codexglue}. 
Please note that all Deep Learning-based approaches directly extract features from commit contents and do not use any hand-made features to augment them. While the main interest of this paper is to explore and identify effective ways to combine expert knowledge with Deep Learning-based approaches.

\vspace{0.1cm}
\noindent\textbf{DeepJIT}~\citep{Hoang2019DeepJITAE}
DeepJIT first represents code change tokens and commit log tokens in fixed-dimension embeddings and uses the convolutional neural network for text (textCNN)~\citep{Kim2014ConvolutionalNN} to automatically extract features from the embeddings. In addition, DeepJIT proposes a framework to hierarchically encode structures in commits (i.e. line→hunk→file→commit).

\vspace{0.1cm}
\noindent\textbf{CC2Vec}~\citep{Hoang2020CC2VecDR}
CC2Vec learns the distributed representation of code changes guided by their commit logs. During training, CC2Vec learns code change embeddings by predicting whether a word from the word vocabulary exists in its commit log or not. 
The training objective can help the learned embeddings of code changes to capture the semantic meaning of code changes expressed in commit logs.
\revised{Specifically, CC2Vec uses the Hierarchical Attention Network (HAN)~\citep{DBLP:conf/naacl/YangYDHSH16} model to automatically extract features from the code changes and adopts a multilayer perceptron (MLP) to predict the words in commit logs. HAN can construct vector representations of lines, and hunks, and finally aggregate them into vector representations of commits.}
\revised{When applying CC2Vec to JIT defect prediction,} Hoang et al. integrated the output of CC2Vec with DeepJIT and outperformed the original DeepJIT~\citep{Hoang2020CC2VecDR}.
However, Zeng et al. argued that in their extended JIT defect prediction datasets, CC2Vec could not outperform DeepJIT consistently~\citep{Zeng2021DeepJD}.

\vspace{0.1cm}
\noindent\textbf{CodeBERT}~\citep{codebert} is a bi-modal pre-trained model, which is capable of modeling both natural languages (NL) and programming languages (PL). 
CodeBERT leverages a 12-layer Transformer encoder as the model and pre-trained the encoder with the NL-PL data pairs from the CodeSearchNet dataset~\citep{codesearchnet}.
It adopts two pre-training objectives jointly: Masked Language Modeling (MLM)~\citep{bert} and Replaced Token Detection (RTD)~\citep{electra}. 
The eventual loss function for CodeBERT is formulated below:
\begin{equation}
    \underset{\theta}{\text{min}}( 
    \mathcal{L}_{MLM}(\theta)+\mathcal{L}_{RTD}(\theta))
\end{equation} 
where $\theta$ denotes the model parameters of the 12-layer Transformer encoder.

\revised{As previous works do not include CodeBERT as the baseline, we need to tune the hyperparameters of CodeBERT to reach the best performance. We did the hyper-parameter tuning for CodeBERT by considering the following important hyper-parameters:}

\begin{itemize}
    \item \revised{\textit{Dropout Rate}: Dropout is a regularization technique used in neural networks. The dropout rate is a hyperparameter that determines the probability of randomly setting a fraction of the input units (neurons) to zero during each update during training. It helps prevent overfitting by reducing the reliance of the model on any individual neuron and encourages the network to learn more robust features. We did a grid search on the potential values \textit{dropout-rate=[0.0, 0.2, 0.5, 0.8]}.}
    \item \revised{\textit{Learning Rate}: Learning rate is a crucial hyperparameter that controls the step size at which a model's weights are updated during training. It determines how quickly or slowly a model learns and converges to an optimal solution. A high learning rate may cause the model to converge too quickly and potentially overshoot the optimal solution, while a low learning rate may cause slow convergence or the model to get stuck in a suboptimal solution. It is typically set before training and is adjusted during training based on the model's performance. We did a grid search on the potential values \textit{learning-rate=[0.1, 0.01, 0.001, 0.0001, 0.00001]}.}
    \item \revised{\textit{Batch Size}: Batch size refers to the number of training examples used in each iteration (or batch) during the training of a neural network. It is a hyperparameter that impacts both training speed and model generalization. A smaller batch size can provide noisy updates but may help the model generalize better, while a larger batch size can lead to smoother updates but may require more memory and training time. We did a grid search on the potential values \textit{batch-size=[4, 8, 16, 32, 64]}.}
\end{itemize}
\revised{We found the best hyperparameters of CodeBERT are: \textit{dropout-rate=0.5, learning-rate=0.0001, and batch-size=16} based on the best validation performances. We further use the best hyperparameters of CodeBERT when training CodeBERT for the JIT defect prediction}.

\section{Proposed Approach}

\begin{figure*}[t] 
\centering 
\includegraphics[width=\textwidth]{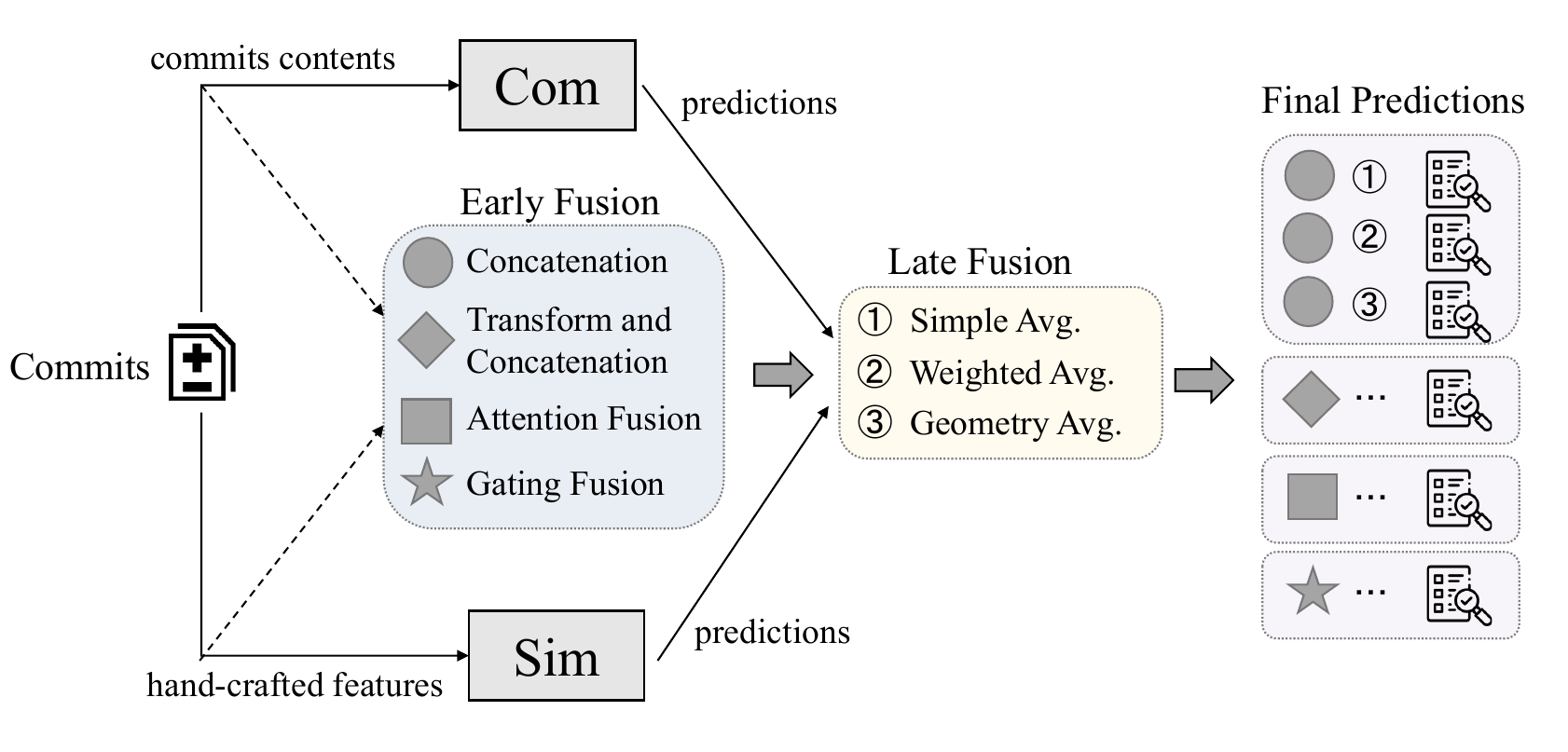}
\caption{The framework to combine simple and complex models. Given a commit, a simple model (e.g., \emph{Sim}) deals with hand-made features based on expert knowledge, and a complex model (e.g., \emph{Com}) extracts features from commit contents. The early fusion strategies can insert expert knowledge into a deep learning model to build another new expert knowledge-enhanced deep learning model. The late fusion strategies combine the prediction scores of the simple model, the complex model, and the model after early fusion, to give final predictions on the defectiveness of commits. }
\label{fusion_combination} 
\end{figure*}

In this section, we present the details of our proposed framework to combine expert knowledge-based models (i.e., simple models) and deep learning-based models (i.e., complex models).
We first propose our own simple model (i.e., \emph{Sim}) and complex model (i.e., \emph{Com}) which can outperform the existing tools.
Then we apply a framework (presented in Figure~\ref{fusion_combination}) to effectively combine the \emph{Sim} and \emph{Com} to achieve better performance. The framework mainly contains three modules: a simple model \emph{Sim}, a complex model \emph{Com}, and two model fusion modules.

\subsection{Simple Model Module: \emph{Sim}}
\label{sec:models}

In this work, we call our simple model module \emph{Sim} for short.
Following the well-known simple model, LR-JIT~\citep{Kamei2013ALE}, we use the same 14 hand-crafted features (as presented in Table~\ref{hand_feature}) in LR-JIT, which describe the properties of commits. 
We choose the features of LR-JIT because LR-JIT considers a comprehensive set of hand-made features based on many prior works~\citep{Mockus2000PredictingRO, Nagappan2006MiningMT,DAmbros2010AnEC, Moser2008ACA, Hassan2009PredictingFU, Koru2009AnII, Guo2010CharacterizingAP, Purushothaman2005TowardUT, Graves2000PredictingFI, Matsumoto2010AnAO}. The effectiveness of the simple model will be affected by the set of hand-made features we choose.

\emph{Sim} utilizes Random Forest (RF)~\citep{Breiman2004RandomF} as the ML classifier, which is one of the most popular ML classifiers. 
RF is a fast and robust classifier that is capable of identifying non-linear patterns in both continuous and categorical features~\citep{Cutler2012}.   
In this model, we directly feed 14 hand-crafted features of commits into the RF and the RF gives predictions on the labels (i.e. defective or clean) of commits. 
We use the implementation of RF in the Scikit-Learn package~\citep{sklearn_api} with the default parameters of the package.

\subsection{Complex Model Module: \emph{Com}}

\begin{figure}[b] 
\centering 
\includegraphics[width=0.9\textwidth]{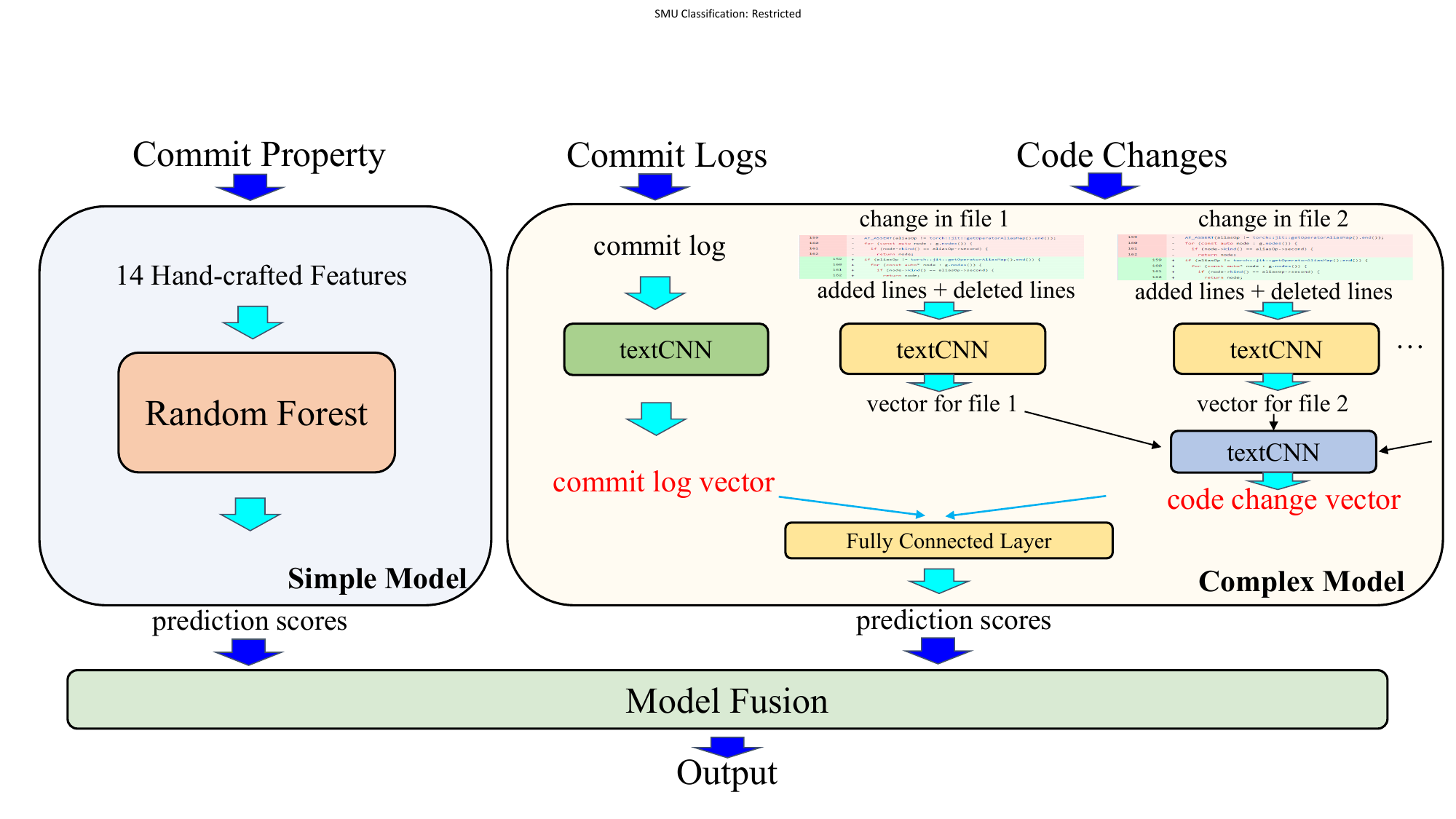}
\caption{Model Details of Com. Com separately encodes and extracts features from code changes and commit logs via different textCNN modules. For code changes, Com considers the hierarchical structure of code changes and extracts features from the line, hunk, file, and finally to the commit level. Com concatenates the commit log vector and code change vector to get the commit vector as the last representation used for prediction.}
\label{com} 
\end{figure}

In this work, we call our complex model module \emph{Com} for short.
Following previous work, in the complex model module, we use textCNN~\citep{Kim2014ConvolutionalNN} as the basic feature extractor to automatically extract features from the contents of commits.
The model details of \emph{Com} is presented in Figure~\ref{com}.
As a commit consists of a commit log and a set of code changes, we first separately extract features from the commit log and code changes respectively (i.e. a commit log vector and a code change vector). 
Then, we concatenate the two vectors to get the final feature for the whole commit (i.e. a commit vector).
After getting the commit vector, we feed the commit vector into a fully connected layer (which acts as a classifier) to predict whether a commit is defective or not. 
In general, this complex model has the similar framework as the prior work DeepJIT.
We follow the framework of DeepJIT since this framework can hierarchically encode structures in commits (i.e. line$\rightarrow$hunk$\rightarrow$file$\rightarrow$commit), which is used in prior DL-based JIT defect predictors~\citep{Hoang2019DeepJITAE, Hoang2020CC2VecDR} and is proven useful.
Though the framework is similar, we make changes to reflect the code changes in a better way.
We will explain the details in Section~\ref{subsection:featureExtraction}.

\subsubsection{\textbf{textCNN}}

The textCNN, as the basic feature extractor, can aggregate a sequence of token embeddings into a single vector that contains the semantic meaning of the input sequence~\citep{Kim2014ConvolutionalNN}.

Given a sequence of tokens denoted as $M = [ t_{1}, t_{2},..., t_{|M|}]$, each token $t_{i}$ has its own embedding $w_{i} \in R^{d}$ where $d$ is the dimension of token embeddings.
$w_{i:j}$ stands for the concatenation of token embeddings from position $i$ to $j$ in the sequence.
To fuse the embeddings of tokens in a sequence, a filter $f \in R^{k \times d}$ is utilized to a window of k files to compute a new feature:
$$ x_{i} = \alpha(f \cdot w_{i:i+k-1} + b_{i})  $$
where $\cdot$ is a sum of element-wise product, $\alpha(.)$ is a non-linear activation function, and $b_{i} \in R$ is the bias value.
The filter $f$ is applied to every k-tokens in a sequence. The generated features are then concatenated to form a vector $X$ such that:
$$X = [x_{1}, x_{2}, ... , x_{|M|-k+1}] $$
Following prior work~\citep{Hoang2019DeepJITAE,Kim2014ConvolutionalNN}, for each filter, we then use a max-pooling layer~\citep{Goodfellow2015DeepL} to process the feature vector $X$ to obtain the highest value:
$$ c = \max_{1 \le i \le |M|-k+1} x_{i}$$
where $c$ is the feature of the input corresponding to a particular filter. 
Therefore, each filter extracts a feature $c$ from the input $M$.
textCNN uses multiple filters (with varying window sizes) to get multiple features~\citep{Kim2014ConvolutionalNN}.
Features from all filters are concatenated to form $Z_{M}$, a single vector that represents the whole sequence. 

In brief, we can simply think of textCNN as a feature  extractor that turns a sequence of tokens into a single vector:
$$ Z_{M} = \mathit{textCNN}(M) $$
where $M$ is the input sequence and $Z_{M}$ is the output feature (i.e. a vector representing the input sequence).

\subsubsection{\textbf{Feature extraction on commit logs}}
\label{subsection:featureExtraction_msg}
Commit logs are mainly natural language (NL) sentences that summarize the purposes of those commits by authors.
As a commit log $m$ is a sequence of NL tokens, textCNN can be used without any pre-processing to extract a single vector from the commit log.
We call this single vector the commit log vector (i.e. $Z_{m}$).

\subsubsection{\textbf{Feature extraction on code changes}}
\label{subsection:featureExtraction}
Unlike a commit log, code changes in a commit are not a simple sequence but a list of sequences: code changes spread over different code files in a project because a commit may modify several files simultaneously.

For changes in each code file, we first concatenate all added and removed lines separately. Then we concatenate the added lines and the removed lines using special headers. 
The concatenated added and removed lines in each file are then processed to be one single sequence to represent changes in a file.
The motivation for this processing comes from the following observation.
In the previous work~\citep{Hoang2019DeepJITAE}, there is no difference in the ways of processing the added code lines and removed code lines (i.e. no special headers as indicators).
We highlight the added (deleted) parts by adding two special headers (i.e., ``Added:'' and ``Removed:'') indicating the property of the following code lines (i.e. added or deleted).
Adding too many special tokens may hurt the model’s performance. To avoid adding special headers to every changed line, we first concatenate the added (deleted) lines together and then only add two special headers.
\revised{Please note that we only add special headers to the changed code lines and the orders and the contexts of added/removed code lines are not changed.}

As the added and removed lines in each file are processed to be one single sequence to represent changes in a file, we then feed the sequence into a textCNN to get the vector for changes in each file.
One important step for effective JIT defect prediction is to find the most suspicious parts of commits. 
This process is very similar to extracting important keywords/phrases from the text to do classification~\citep{Kim2014ConvolutionalNN, zhang2015sensitivity}. textCNN is an efficient deep learning sequence feature extractor, which is good at capturing the key segments of an input text~\citep{zhang2015sensitivity} (e.g. most defective code token sequences in our case).
Thus, we use textCNN to extract features from code changes in a code file.
\revised{Regarding the hyper-parameters of textCNN, we follow previous work DeepJIT~\citep{Hoang2019DeepJITAE} to set the number of filters to 64 and the dimension of word vectors to 64 as well.}

To fuse the vectors (features) from different code files, we use another textCNN to aggregate those vectors into a single vector $Z_{C}$ representing the whole code changes (i.e. the code change vector).

\subsubsection{\textbf{Feature concatenation and prediction}}
\label{sec:z}
After getting the vectors for the commit log and the code changes, we concatenate these two vectors to generate a final feature representation (i.e. $Z$) representing the commit:
$$ Z = Z_{m} \oplus Z_{C} $$
where $\oplus$ is the concatenation operator.

The vector $Z$ is fed into a widely used DL classifier, a fully-connected (FC) layer~\citep{Hoang2019DeepJITAE, Hoang2020CC2VecDR}, to get the prediction score.
$$ y = \text{sigmoid} (\alpha(w_{h} \cdot Z + b_{h}))  $$
where $\cdot$ is a dot product, $w_{h}$ is a weight matrix of the FC layer, $b_{h}$ is the bias term, $\alpha(.)$ is a non-linear activation function, and $\text{sigmoid}(.)$ is a function to do normalization on the prediction scores.
\revised{The prediction score y consists of two elements, i.e., $y = [prob_0, prob_1]$ where $prob_1$ indicates the predicted probability of being defective and the $prob_0$ is the predicted probability of being clean.
The dataset labels are transformed into the same formats as the prediction scores, where a label $l = [0, 1]$ represents a defective instance and $l = [1, 0 ]$ represents a clean instance. 
To update the model parameters, the difference between the prediction score $y = [prob_0, prob_1]$ and the corresponding label $l$  is measured by the widely used Cross-Entropy loss, as follows: 
$$ Loss (y, l) = - \sum y(x) log ( l(x) ) $$
where x refers to 0 and 1 respectively. 
Finally, the whole complex model is updated together to minimize the Cross-Entropy loss.
}

\subsection{Model Fusion}

JIT defect prediction can be regarded as a multi-modal task:  hand-crafted features, commit logs, and code changes can be seen as three different aspects (modalities) of commits.  These three modalities have different properties: hand-crafted features are numerical (or categorical) features with clear meanings; commit logs are natural language texts, and the code changes are written in a programming language. 
Existing JIT defect prediction approaches only use either expert knowledge or the commit contents (commit logs and code changes) to assess the defectiveness of a commit. However, these two sources of information both matter as they may present different aspects of the commits. In absence of multi-modal features, existing JIT defect prediction approaches may be prone to make wrong predictions. 
In this paper, we explore ways to insert expert knowledge into the DL-based approaches and obtain more accurate predictions.

In general, model fusion techniques are widely used in tasks that have multiple data modalities such as object detection and video analysis~\citep{Snoek2005EarlyVL,Gunes2005AffectRF,Tsekeridou2001ContentbasedVP,Ma2019InfraredAV,Dong2013PerformanceEO} where multi-modal data exist. There are two typical fusion methods, namely early fusion and late fusion.
Early fusion is carried out at the feature level: various features of different modalities are combined into one vector for classification~\citep{Snoek2005EarlyVL,Gunes2005AffectRF}.
Late fusion (also called decision level fusion) is to build several models independently for different modalities (or modality groups) and do a fusion at a decision-making stage (on the prediction scores)~\citep{Snoek2005EarlyVL,Gunes2005AffectRF}.

\begin{figure*}[thbp]
    \centering
    \includegraphics[width=0.35\textwidth]{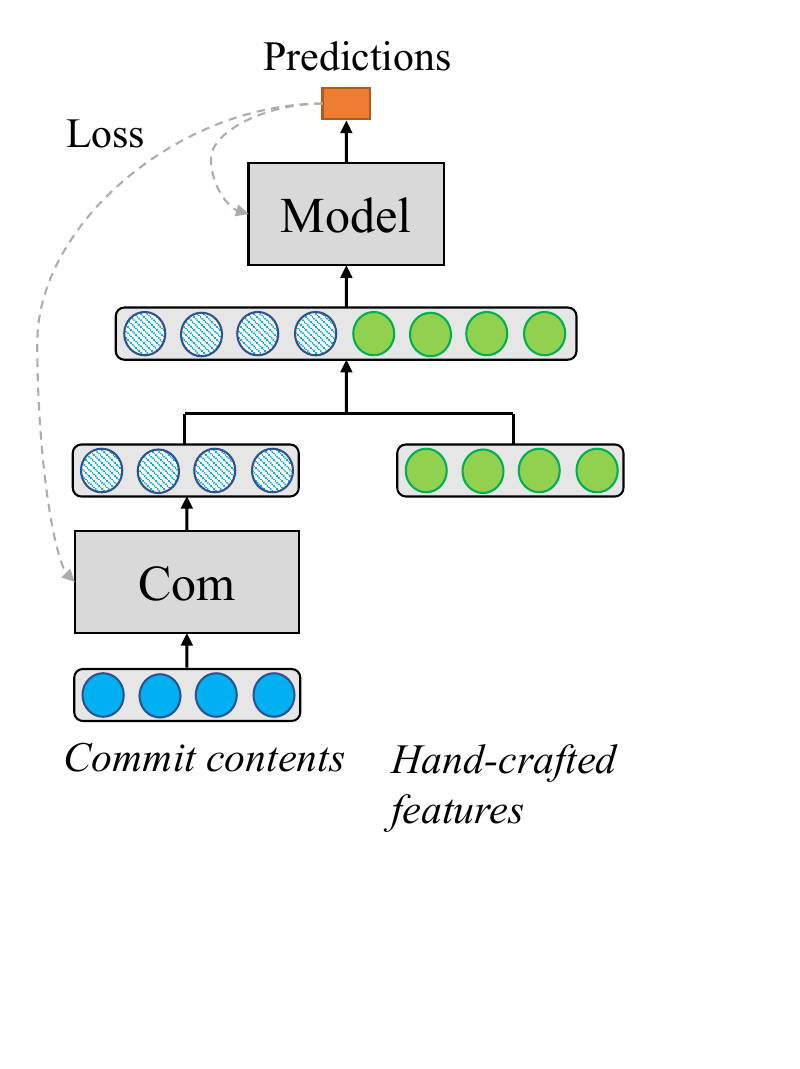}
    \includegraphics[width=0.45\textwidth]{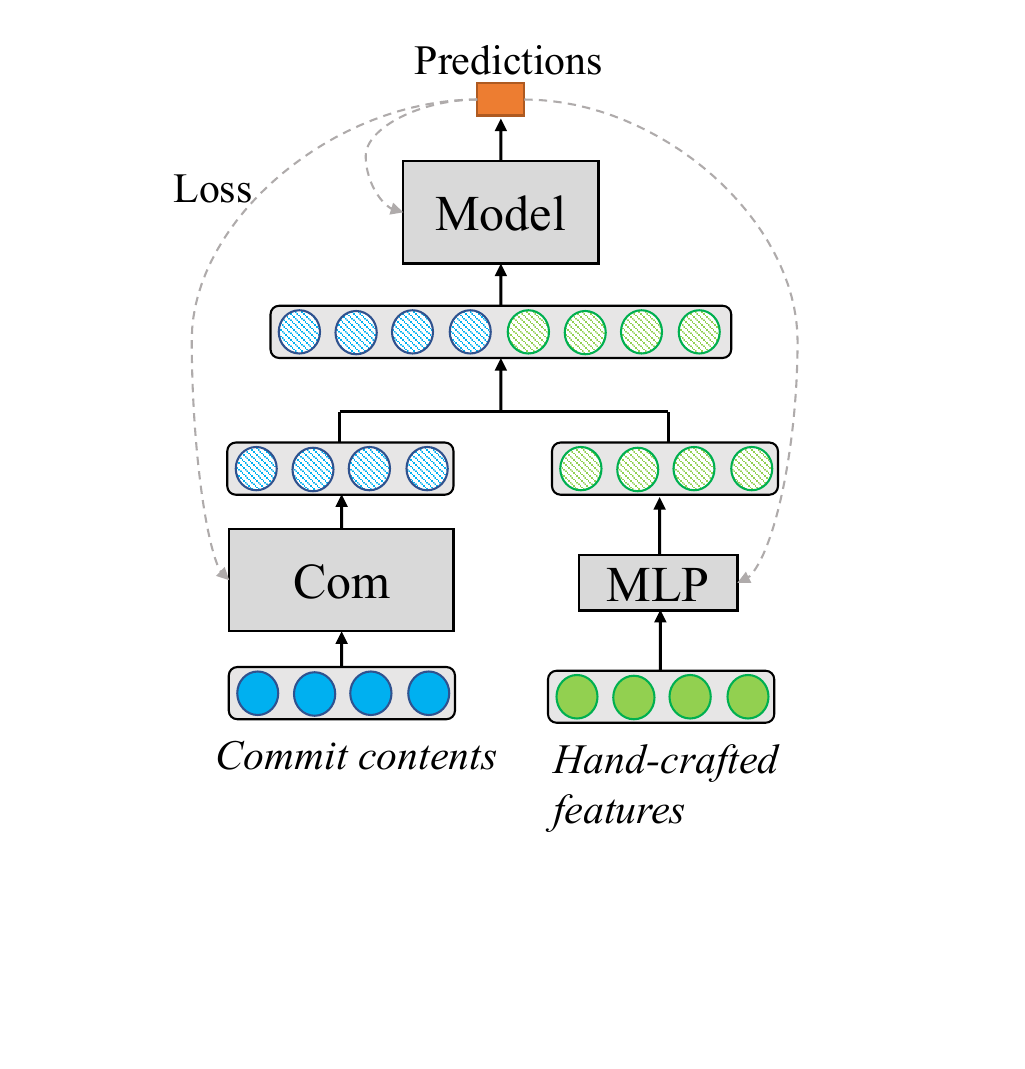}
    \\
    \makebox[0.35\textwidth]{\small  (a) Simple Concatenation} 
    \makebox[0.45\textwidth]{\small (b) Transform and Concatenation}
    \\ \vspace{0.5em}
    \includegraphics[width=0.4\textwidth]{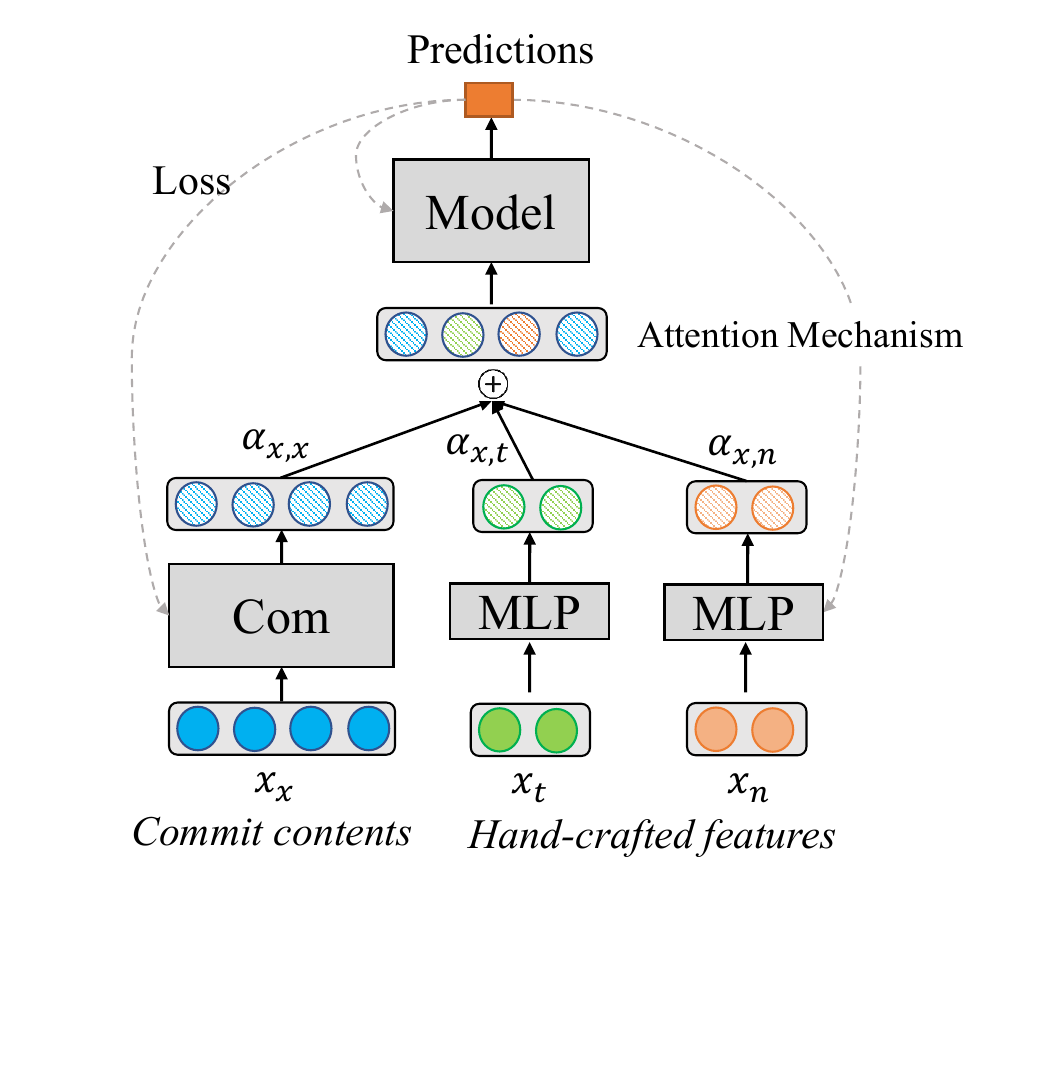}
    \includegraphics[width=0.4\textwidth]{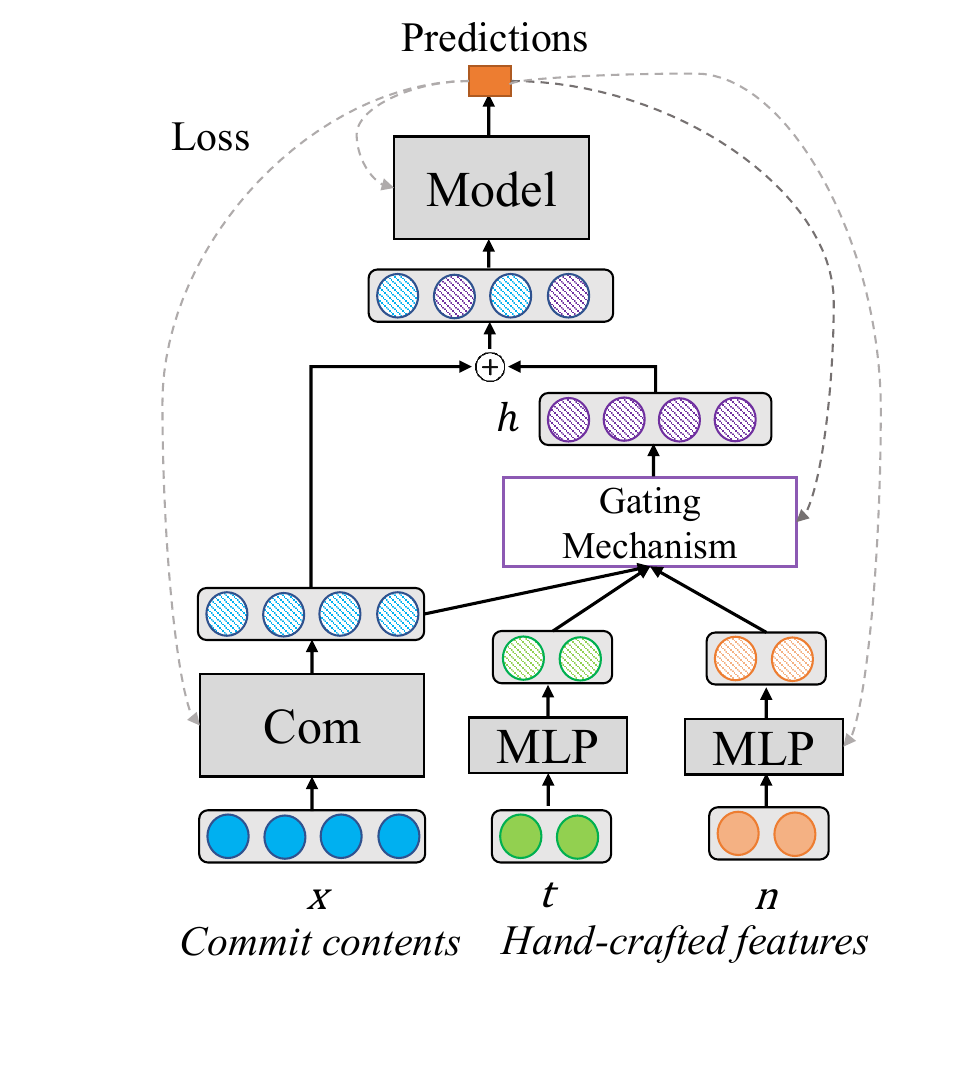}
     \\
    \makebox[0.4\textwidth]{\small (c) Attention Mechanism Fusion}
    \makebox[0.4\textwidth]{\small (d) Gating Mechanism Fusion}
    \caption{Visualization of Early Fusion Techniques.}
    \label{early_fusion}
\end{figure*}

\subsubsection{\textbf{Early Fusion}}

Early fusion strategies combine various features of different modalities into one vector and the combined vectors are input into a fully connected (FC) layer to predict the labels. 
In this study, we identify four different features to fuse:
\begin{enumerate}[]
\item Commit log vectors ($\mathbf{ Z_{m}}$): they are the features automatically extracted from the commit message (natural language texts).

\item Code change vector ($\mathbf{ Z_{c}}$): they are the features automatically extracted from the code changes (programming language codes).

\item Categorical hand-crafted feature ($\mathbf{X_{cat}}$): they are hand-crafted features (expert knowledge) whose values are discrete.

\item Numerical hand-crafted feature ($\mathbf{X_{cont}}$): they are hand-crafted features (expert knowledge) whose values are continuous.

\end{enumerate}

Given features from different modalities above, we study the following different strategies to carry out early fusion (feature-level fusion).

\vspace{0.2cm}
\noindent\textbf{Simple Concatenation (SC).} It simply concatenates various features of different modalities into a single vector. Formally, we have
$$\mathbf{C} = \mathbf{Z_m} \, \oplus \, \mathbf{Z_c} \, \oplus \, \mathbf{X_{cat}} \, \oplus \, \mathbf{X_{cont}}$$ where $\mathbf{C}$ is the combined vector after the early fusion module and $\oplus$ is the concatenation operation.

\vspace{0.2cm}
\noindent\textbf{Transform and Concatenation (TC).} 
In the simple concatenation strategy, the hand-crafted features are fixed and may not be compatible with features automatically extracted from commit contents that are updated during training. 
To mitigate this issue, this strategy first performs the matrix transformation on the categorical and numerical hand-crafted features. Thus, it transforms the fixed hand-crafted features into learnable features because the matrix can be updated during training.
Then, it concatenates the transformed hand-made features with the features of commit logs and code changes.
$$ \mathbf{C} = \mathbf{Z_m} \, \oplus \, \mathbf{Z_c} \, \oplus \, (\mathbf{W_{cat}} \odot \mathbf{X_{cat}}) \, \oplus \, (\mathbf{W_{cont}} \odot \mathbf{X_{cont}}) \, $$
$\mathbf{W_{cat}} \in R^{dm \times dm}$ is the matrix for categorical features where $dm$ is the dimension of $\mathbf{X_{cat}}$. 
$\mathbf{W_{cont}} \in R^{k \times k}$ is a matrix for $\mathbf{X_{cont}}$ and $k$ is the dimension of numerical features. 
In addition, ``$\odot$'' refers to matrix multiplication.
Please note that $\mathbf{W_{cat}} \odot \mathbf{X_{cat}}$ (or $\mathbf{W_{cont}} \odot \mathbf{X_{cont}}$)  have the same sizes as $\mathbf{X_{cat}}$ (or $\mathbf{X_{cont}}$). 
$\mathbf{W_{cat}}$ and $\mathbf{W_{cont}}$ are two learnable matrices and their values are updated by  minimizing the binary cross-entropy.

\vspace{0.2cm}
\noindent\noindent\textbf{Attention Mechanism Fusion (AMF).}
In the prior two early fusion strategies, the features from different modalities are directly concatenated but lack cross interactions between modalities. Attention mechanism~\citep{vaswani2017attention,bahdanau2014neural} is an effective technique to pay different weights to different regions in each modality and include the correlation between commit contents and hand-crafted features. In this strategy, we apply an attention mechanism between the commit content features (both commit logs and code changes ) and hand-crafted features (both categorical and numerical features)  to achieve better performance. 
For ease of explaining the cross-attention feature fusion, we simplify features into the following vectors:
$$ \mathbf{x_x} = \mathbf{Z_m} \, \oplus \, \mathbf{Z_c} \, $$
$$ \mathbf{x_t} = \mathbf{X_{cat}} $$
$$ \mathbf{x_n} = \mathbf{X_{cont}} $$
Formally, the combined vector is expressed as a weighted average:
$$ \mathbf{C} = \alpha_{x,x}\mathbf{W}_x\mathbf{x_x} + \alpha_{x,t}\mathbf{W}_t\mathbf{x_t} + \alpha_{x,n}\mathbf{W}_n\mathbf{x_n} $$
where $\mathbf{W}_x \in R^{dim \times dim}$ and $dim$ is the dimension of $\mathbf{x_x}$.
$\mathbf{W}_t \in R^{dm \times dm}$ and $dm$ is the dimension of $\mathbf{x_t}$.
$\mathbf{W}_n \in R^{k \times k}$ and $k$ is the dimension of $\mathbf{x_n}$. 
Besides, the attention coefficients$\alpha_{i,j}$ are computed as:
$$  \alpha_{i,j} =
                \frac{
                \exp\left(\mathrm{LeakyReLU}\left(\mathbf{a}^{\top}
                [\mathbf{W}_i\mathbf{x}_i \, \oplus \, \mathbf{W}_j\mathbf{x}_j]
                \right)\right)}
                {\sum_{k \in \{ x, t, n \}}
                \exp\left(\mathrm{LeakyReLU}\left(\mathbf{a}^{\top}
                [\mathbf{W}_i\mathbf{x}_i \, \oplus \, \mathbf{W}_k\mathbf{x}_k]
                \right)\right)}. $$
where $\mathrm{LeakyReLU}$~\citep{radford2015unsupervised} is an activation function and $\mathbf{a}$ is the attention vector. \revised{We choose to use $\mathrm{LeakyReLU}$ as the activation function because it is computationally efficient and resilient against the vanishing gradient problem. It can also address the "dead neuron" issue, where neurons with negative biases may never be activated~\citep{radford2015unsupervised}.}

\vspace{0.2cm}
\noindent\textbf{Gating Mechanism Fusion (GMF).}
The gating mechanism is widely used in neural networks~\citep{gru,dauphin2017language} to regulate the information to be kept or discarded at each step.
Here we use this mechanism to highlight the relevant information in expert knowledge and commit content features while discarding that irrelevant information by gates. In this way, the learning ability of deep learning modules can focus on learning the relevant information.
For ease of presentation, we still simplify the following vectors:
$$ \mathbf{x} = \mathbf{Z_m} \, \oplus \, \mathbf{Z_c} \, $$
$$ \mathbf{t} = \mathbf{X_{cat}} $$
$$ \mathbf{n} = \mathbf{X_{cont}} $$
As shown above, a GMF unit receives three inputs, one is the commit contents, one is categorical features, and the last one is numerical features. 
We create bimodal pairs [$x$;$t$] and [$x$; $n$] by concatenating the commit content vector with categorical and numerical features respectively.
The bimodal pairs are used to generate two gating vectors $g_t$ and $g_n$, \revised{following previous works in the machine learning area which adopted Gating Mechanism~\citep{wang2019words,zhang2016gated}.}
$$\mathbf{g}_t = R(\mathbf{W}_{gt}[\mathbf{t} \, \oplus \, \mathbf{x}]+ b_t)$$
$$\mathbf{g}_n = R(\mathbf{W}_{gn}[\mathbf{n} \, \oplus \, \mathbf{x}]+ b_n)$$
These gates highlight the relevant information between categorical (numerical) features and the commit content vector.
We then create a vector $h$ that represents the expert knowledge information after gates by fusing together $t$ and $n$ multiplied by their respective gating vectors:
$$ \mathbf{h} = \mathbf{g_t} \odot (\mathbf{W}_t\mathbf{t}) + \mathbf{g_n} \odot (\mathbf{W}_n\mathbf{n}) + b_h $$
 \revised{The gates ($\mathbf{g}_t$ and $\mathbf{g}_n$) allow the network to selectively pass information and update weights. For example, information from $\mathbf{t}$ to $\mathbf{h}$ is controlled by the $\mathbf{g}_t$.}

Finally, we use a weighted summation between the commit content vector $x$ and the expert knowledge vector $h$
to create a multi-modal vector $C$:
$$  \mathbf{C}= \mathbf{x} + \alpha\mathbf{h} $$
\revised{The scales of $\mathbf{h}$ and $\mathbf{x}$ may vary significantly. When the scale of $\mathbf{h}$ is considerably larger than that of $\mathbf{x}$ (e.g.,  $\frac{\| \mathbf{h} \|_2}{\| \mathbf{x} \|_2}=100$), the neural network tends to prioritize optimizing $\mathbf{h}$ over $\mathbf{x}$. However, our objective is for the neural network to perform well for both x and h. To achieve this balance, we introduce a weight factor $\alpha$ to rescale the magnitude of $\mathbf{h}$. We follow a previous work~\citep{wang2019words} in machine learning to calculate the weight factor $\alpha$ by:}
$$ \alpha = \mathrm{min}( \frac{\| \mathbf{x} \|_2}{\| \mathbf{h} \|_2}*\beta, 1) $$
where $\beta$ is a hyperparameter selected through the validation process. $\| \mathbf{x} \|_2$ and $\| \mathbf{h} \|_2$ denote the L2 norm~\citep{cortes2012l2} of the $x$ and $h$ vectors respectively.

\begin{figure}[t] 
\centering 
\includegraphics[width=0.5\textwidth]{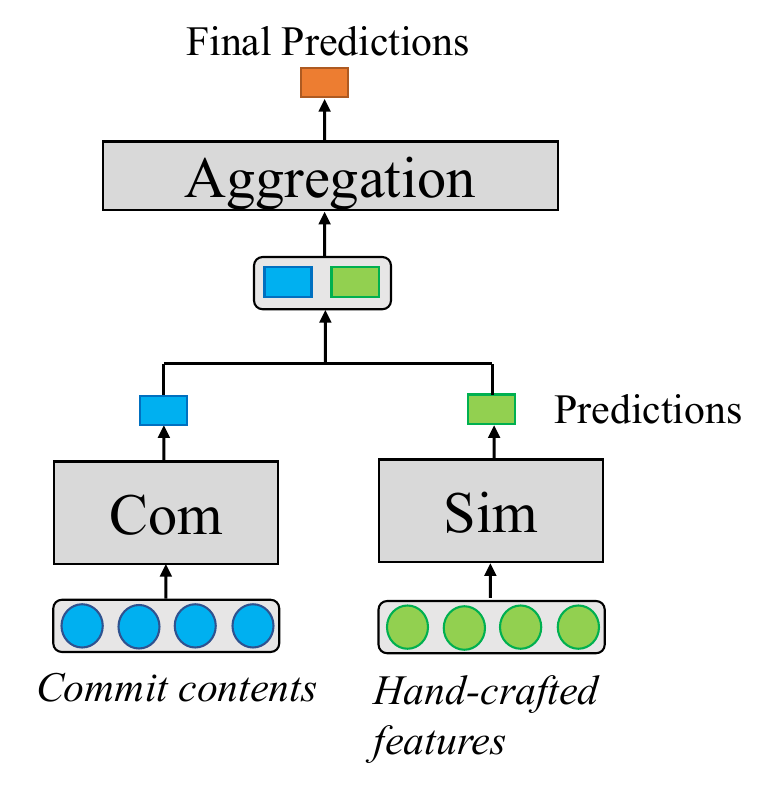}
\caption{Example of Late Fusion Strategies. }
\label{late_fusion} 
\end{figure}

\subsubsection{\textbf{Late Fusion}}
Late fusion is to fuse decisions of different models~\citep{Snoek2005EarlyVL,Gunes2005AffectRF}. 
We first build several models independently for different modalities (or modality groups).
Then, we give final predictions by voting predictions of different models. 
Specifically, late fusion has different rules to determine how to finally combine each of the outputs of the independently trained models, e.g. product, sum, average, or weighted average of the outputs of all models~\citep{Yu2010L2normMK}.
For instance, if we only have two models (i.e., \emph{Sim} and \emph{Com}) for expert knowledge and commit contents respectively, we could train \emph{Sim} and \emph{Com} separately, and each model will give prediction scores for commits. 
As shown in Figure~\ref{late_fusion}, these scores could be combined in a late fusion fashion to yield a final prediction score. 
In a generalized case, given $n$ separate models, the $i$-th model gives a prediction score on the $j$-th patch denoted as $p_{j}^{i}$. The final predictions on the $j$-th patch using different strategies are shown below: 

\vspace{0.2cm}
\noindent\textbf{Simple Average.} The final prediction on the $j$-th patch is:
    $$  p_j = \frac{ \sum_{i=1}^{n} p_{j}^{i} }{n}  $$

\vspace{0.2cm}
\noindent\textbf{Weighted Average.} This strategy needs to define weights of models w=[$w_1$, $w_2$, ..., $w_n$]. The weights can be either optimized during validation or arbitrarily chosen beforehand.
    The final prediction on the $j$-th patch in this late fusion strategy is:
    $$  p_j = \frac{ \sum_{i=1}^{n}  p_{j}^{i}*w_i }{ \sum_{i=1}^{n} w_i}  $$

\vspace{0.2cm}
\noindent\textbf{Geometric Average.}  It is an average that indicates a central tendency of a set of numbers by using the product of their values. The final prediction on the $j$-th patch in this late fusion strategy  is:
    $$  p_j =  \sqrt[1/n] {\prod_{i=1}^{n}  p_{j}^{i} }  $$

\subsubsection{\textbf{Early/Late Fusion Combination}}

 In the model training phase, we can train one simple model (i.e. \emph{Sim}) and one complex model (i.e. \emph{Com}). Also, we can train a fused model using early fusion strategies.
During validating or testing, each model will give a prediction score for a commit and these scores are aggregated via late fusion strategies to yield a final prediction score.
There are 5 variants in early fusion (SC, TC, AMF, GMF, and None) and 4 variants in late fusion (Simple Average, Weighted Average, Geometric Average, and None) where ``None'' indicates that does not use an early/late fusion strategy. In this case, we experiment with 20 fusion combinations to explore and identify the fused best model for the JIT defect prediction.
Please note that we choose the best-performing combination during validation as the final proposed model.

\section{Experimental Setting}

\subsection{Datasets}
In this paper, we select the datasets collected by Zeng et al.~\citep{Zeng2021DeepJD} for the following reasons.
First, to make a fair comparison, we choose to use the same training and testing datasets with the state-of-the-art~\citep{Zeng2021DeepJD}. 
Second, we would like to carry out experiments on high-quality JIT defect prediction datasets.
JIT defect prediction datasets are mainly collected by utilizing the SZZ algorithm~\citep{Kamei2013ALE, Hoang2019DeepJITAE, Yan2020JustInTimeDI}. 
When collecting the datasets, Zeng et al. paid attention to two known issues of the SZZ algorithm (i.e. the false label issue~\citep{Kim2006AutomaticIO, McIntosh2018AreFC} and authentication latency issue~\citep{Tan2015OnlineDP, Cabral2019ClassIE}) and followed prior works~\citep{Kim2006AutomaticIO, McIntosh2018AreFC, Cabral2019ClassIE} to reduce the impacts of the known issues.
Finally, the dataset of Zeng et al. 
is a diverse set of real-world projects. They collect all code commits in six popular projects over the last 10 years.

Table~\ref{data_stat} shows the statistics of six studied datasets including the project name and the number of commits in the training and test set respectively.
For the columns, the ``Defect'' presents the number of commits that are defective in the dataset.
The ``Clean'' column stands for the number of commits that are clean (not defective).

\revised{Although we also want to do experiments with the dataset used in the DeepJIT paper~\citep{Hoang2019DeepJITAE}, the official replication package of DeepJIT\footnote{https://github.com/soarsmu/DeepJIT} only contains the commit contents (for training Deep Learning-based approaches). The hand-crafted features (for traditional Machine Learning-based approaches) are not available. As DeepJIT is a work done before 2019, it is hard for the authors of DeepJIT to collect the hand-crafted features because necessary intermediary files have already been lost. Hence, we have excluded the dataset from the DeepJIT paper in our study.
}

\begin{table}[!t]
\caption{Datasets statistics}
\begin{tabular}{|l|cc|cc|}
\hline
\multirow{2}{*}{\textbf{Statistics}} & \multicolumn{2}{c|}{\textbf{Training Set}}            & \multicolumn{2}{c|}{\textbf{Testing Set}}             \\ \cline{2-5} 
                                     & \multicolumn{1}{c|}{\textbf{Clean}} & \textbf{Defect} & \multicolumn{1}{c|}{\textbf{Clean}} & \textbf{Defect} \\ \hline
\textbf{QT}                          & \multicolumn{1}{c|}{16,242}          & 2,887            & \multicolumn{1}{c|}{4,088}           & 695             \\ \hline
\textbf{OpenStack}                   & \multicolumn{1}{c|}{13,153}          & 5,052            & \multicolumn{1}{c|}{3,677}           & 875             \\ \hline
\textbf{JDT}                         & \multicolumn{1}{c|}{1,342}           & 1,281            & \multicolumn{1}{c|}{335}            & 321             \\ \hline
\textbf{Platform}                    & \multicolumn{1}{c|}{5,477}           & 3,350            & \multicolumn{1}{c|}{1,501}           & 706             \\ \hline
\textbf{Gerrit}                      & \multicolumn{1}{c|}{1,0348}          & 1,593            & \multicolumn{1}{c|}{2798}           & 188             \\ \hline
\textbf{GO}                          & \multicolumn{1}{c|}{8,530}           & 6,677            & \multicolumn{1}{c|}{2,334}           & 1,468            \\ \hline
\end{tabular}
\label{data_stat}
\end{table}

\subsection{Evaluation metrics}
We use three widely-used evaluation metrics to evaluate the performance of our approach and baselines.

\textbf{AUC-ROC}:
Receiver Operator Characteristic (ROC) curves are used in the binary classification tasks, which show how the number of correctly classified positive examples varies with the number of wrongly classified negative examples~\citep{Provost1998TheCA}.
The Area Under the ROC Curves (AUC-ROC) is commonly used to present the performance of approaches in binary classification tasks in software engineering~\citep{Misirli2009ReducingFA,Rana2009AnFF,Lessmann2008BenchmarkingCM,Menzies2007DataMS,Menzies2008ImplicationsOC,Hoang2019DeepJITAE,Hoang2020CC2VecDR,Zeng2021DeepJD}.
Following the previous work in JIT defect prediction~\citep{Hoang2019DeepJITAE, Hoang2020CC2VecDR, Zeng2021DeepJD}, we adopt this widely-used AUC-ROC score to evaluate the performance of models. 
Besides, the AUC-ROC score does not need one to manually set a threshold~\citep{Tantithamthavorn2020TheIO}. 

\revised{While AUC-ROC has traditionally been the primary evaluation metric in prior JIT defect prediction research, it is essential to acknowledge that it can sometimes provide an overly optimistic assessment of model performance when dealing with imbalanced datasets~\citep{Davis2006TheRB,branco2016survey}.  In such cases, AUC-PR emerges as a robust alternative, widely employed in the field of information retrieval research~\citep{Davis2006TheRB}. 
Given the presence of highly imbalanced datasets in our study, we have chosen to also utilize the AUC-PR metric.
}

\textbf{AUC-PR}:
Precision-Recall (PR) curves, often used in information retrieval~\citep{Manning2002FoundationsOS}, have been used as an alternative to ROC curves for tasks with a large skew in the class distribution\citep{Goadrich2004LearningEO,Davis2005ViewLF,Bunescu2005ComparativeEO,Bockhorst2004MarkovNF,Davis2006TheRB}.
Besides, Saito and Rehmsmeier found that PR curves are more informative than ROC curves when evaluating binary classifiers on imbalanced datasets~\citep{Saito2015ThePP}.
As shown in Table~\ref{data_stat}, all of the six datasets are imbalanced datasets.
For instance, in the test set of Gerrit, the number of clean commits is 13.88 times bigger than the number of defective commits.
Thus, we adopt the Area Under the PR Curves (AUC-PR) as an evaluation metric in this study.
Please note that the AUC-PR score also does not need to set a threshold manually.

\textbf{F1 scores}:
In this work, we also consider the F1 score as the metric. 
F1-score is a popular metric used in many prior software engineering studies~\citep{Jiang2013PersonalizedDP,Arisholm2007DataMT,Rahman2012RecallingT,Shihab2012StudyingRB,Garcia2018CharacterizingAP}, and is defined as the harmonic mean of precision and recall:
F1-score =$ \frac{2\times Recall \times Precision}{Recall + Precision}.$
Precision and recall, in turn, are defined based on the number of true positives (TP), false positives (FP), and false negatives (FN).
If a defective commit is classified as defective, it is a true positive (TP), and otherwise false negative (FN).
If a clean commit is mistakenly classified as defective, it is a false positive (FP).
Precision is the ratio of correctly predicted defective commits to all commits predicted as defective
 (i.e. Precision~$=\frac{TP}{TP + FP}$).
Recall is the ratio of the number of correctly predicted defective commits to the actual number of defective commits (i.e. Recall~$=\frac{TP}{TP + FN}$).

\subsection{Train-Test Pipelines}

We follow the same data partitioning of the LAPredict work~\citep{Zeng2021DeepJD}, i.e. the earlier 80\% data are the training set and the later 20\% data are the testing set.
We use the same data split for all methods, including the three baseline methods.
As our complex model is a DL-based model which has many hyper-parameters, we set aside 5\% of data from training data as the fixed validation set to tune the hyper-parameters of the complex model.
Therefore, the data split ratio for SimCom++ is 75\%, 5\%, and 20\% for training, validation, and testing respectively.
We use the original hyper-parameters for the baselines. 
Please note that all approaches are evaluated on the same hold-out test sets in a manner of fair comparison. 
For the training of baselines, we use scripts in the replication packages and use the default hyper-parameters in the packages.

For SimCom++, we separately train our simple model (i.e. \textit{Sim}), our complex model (i.e. \textit{Com}), and the deep learning model after early fusion with expert knowledge.  
For training of the simple model (i.e. \emph{Sim}), we use the default parameters of Random Forest (RF) in the Scikit-Learn package.
As shown in Table~\ref{data_stat}, the experiment datasets are imbalanced: the number of defective commits is smaller than clean commits.
The imbalance may cause performance degradation of the ML models if it is not handled properly~\citep{Kamei2013ALE, Kamei2007TheEO, Khoshgoftaar2004BalancingMR}.
To mitigate the class imbalance issue for our simple model, we follow the prior work~\citep{Kamei2013ALE} to use an undersampling approach for our training data.
We randomly delete the majority class instances (i.e. clean commits in training data) until the majority class drops to the same level as the minority class (i.e. defective commits).
Please note that we do not undersample the test data.

For training of the complex model (i.e. \emph{Com}) and the deep learning model after early fusion, we set the learning rate as $5e^{-5}$ and the batch size as 64 after tuning them in the validation.
For other hyper-parameters, we set the size of the fully-connected layer as 512, the number of epochs as 30, and the dropout rate as 0.5.
These hyper-parameter settings are commonly used in prior work~\citep{Severyn2015LearningTR, Hoang2019DeepJITAE,  Huo2016LearningUF, Hinton2012ImprovingNN}.
We use the Adam optimizer~\citep{Kingma2015AdamAM} to update model parameters.
To address the class imbalance issue for our complex model, we use the same loss function as the previous work~\citep{Hoang2019DeepJITAE}.

For validation, we tune the hyper-parameters of our deep learning models by observing the performance of the trained model on the validation set. 
Specifically,
we tune the hyper-parameters using a grid search procedure
with the following set of parameters and their possible values:
the learning rate is in 
$\{1e^{-5}, 5e^{-5}, 1e^{-4}, 2e^{-4}\}$ and the batch size is in $\{16, 32, 64, 128\}$.
We select the combination of hyper-parameters that lead to the best performance on the validation set.
The best-performing model in the validation is used for the evaluation of the test sets.
For our simple model, we directly use the default parameters of Random Forest and do not tune them.
For testing, the evaluation is carried out on a hold-out test set which is the test set of each dataset.

\subsection{Research Questions}

{\bf RQ1:  How effective are expert knowledge-based and deep learning-based JIT defect prediction techniques \revised{prior to fusion}?}
\revised{This RQ aims to find the best-performing expert knowledge-based and deep learning-based tools respectively.}
To answer this RQ,  we carry out experiments by adopting typical JIT defect prediction baselines. Those approaches either only use expert knowledge (hand-crafted features) or only use deep learning techniques to extract features from commit contents (code changes and commit logs). In addition, we also include a popular pre-trained model of code (i.e., CodeBERT) because it has shown its great effectiveness in a wide range of SE tasks~\citep{zhou2021assessing}.
We also experiment with our proposed component models: \emph{Sim} and \emph{Com}.

{\bf RQ2:   How different are the predictions of expert knowledge-based and deep learning-based approaches?}
In this RQ, we aim to explore the difference between the predictions of expert knowledge-based tools and deep learning-based tools. 
Different predictions may indicate that models have
learned different aspects of commits. Thus, it is more likely to achieve improvements by combining them.
To answer this RQ, we first visualize the differences in the predictions of expert knowledge-based and deep learning-based approaches by Venn Diagrams. Then, we carry out Wilcoxon signed-rank tests to test whether the predictions of expert knowledge-based and deep learning-based approaches are significantly different from each other.

{\bf RQ3:  What are the best model fusion strategies in the JIT defect prediction?}
To answer this RQ, we explore the best model combination using the framework proposed in this paper. In general, we select the early and late fusion during validation. \revised{Subsequently, we identify the best model fusion combination to form our SimCom++ model. Finally, we assess the performance of SimCom++ in comparison to the baseline models.}

{\bf RQ4:  \revised{Why does the model fusion lead to better effectiveness?}}
In this RQ, we aim to explain \revised{why} the model fusion works in the JIT defect prediction. 
Firstly, we visualize the prediction results of Sim, Com, and the model after model fusion. 
Secondly, we analyze to what extent the model fusion can help to ``correct'' the wrong predictions of its component models. 
Lastly, we present case studies to demonstrate how expert knowledge helps deep learning-based models.

\section{Experimental Results}

\revised{In this section, we perform experiments to answer RQ1 and RQ3, and we delve into case studies to provide insights into RQ2 and RQ4.}

\subsection{\textbf{RQ1: How effective are expert knowledge-based and deep learning-based JIT defect prediction techniques \revised{prior to fusion}?}}

\begin{table}[]
 \centering
\caption{The performance of baselines and our Sim and Com models with respect to AUC-ROC, AUC-PR, and F1 scores}
\resizebox{\columnwidth}{!}{%
\begin{tabular}{|l|l|r|r|r|r|r|r|r|}
\hline
\textbf{Tools}                    & \multicolumn{1}{c|}{\textbf{Metric}} & \multicolumn{1}{c|}{\textbf{QT}}       & \multicolumn{1}{c|}{\textbf{OP.}}      & \multicolumn{1}{c|}{\textbf{JDT}}      & \multicolumn{1}{c|}{\textbf{Plat.}}    & \multicolumn{1}{c|}{\textbf{Gerrit}}   & \multicolumn{1}{c|}{\textbf{GO}}       & \multicolumn{1}{c|}{\textbf{Avg.}}     \\ \hline
                                     & \cellcolor[HTML]{FFFFFF}ROC          & \cellcolor[HTML]{FFFFFF}0.694          & \cellcolor[HTML]{FFFFFF}0.742          & \cellcolor[HTML]{FFFFFF}0.682          & \cellcolor[HTML]{FFFFFF}0.646          & \cellcolor[HTML]{FFFFFF}0.682          & \cellcolor[HTML]{FFFFFF}0.675          & \cellcolor[HTML]{FFFFFF}0.687          \\ \cline{2-9} 
                                     & \cellcolor[HTML]{EFEFEF}PR           & \cellcolor[HTML]{EFEFEF}0.289          & \cellcolor[HTML]{EFEFEF}0.437          & \cellcolor[HTML]{EFEFEF}0.655          & \cellcolor[HTML]{EFEFEF}0.439          & \cellcolor[HTML]{EFEFEF}0.174          & \cellcolor[HTML]{EFEFEF}0.546          & \cellcolor[HTML]{EFEFEF}0.423          \\ \cline{2-9} 
\multirow{-3}{*}{\textbf{LR-JIT}}    & \cellcolor[HTML]{CBCEFB}F1           & \cellcolor[HTML]{CBCEFB}0.337          & \cellcolor[HTML]{CBCEFB}0.430          & \cellcolor[HTML]{CBCEFB}0.623          & \cellcolor[HTML]{CBCEFB}0.450          & \cellcolor[HTML]{CBCEFB}0.158          & \cellcolor[HTML]{CBCEFB}0.558          & \cellcolor[HTML]{CBCEFB}0.426          \\ \hline
                                     & ROC                                  & 0.666                                  & 0.729                                  & 0.587                                  & 0.697                                  & 0.671                                  & 0.674                                  & 0.671                                  \\ \cline{2-9} 
                                     & \cellcolor[HTML]{EFEFEF}PR           & \cellcolor[HTML]{EFEFEF}0.249          & \cellcolor[HTML]{EFEFEF}0.412          & \cellcolor[HTML]{EFEFEF}0.584          & \cellcolor[HTML]{EFEFEF}0.506          & \cellcolor[HTML]{EFEFEF}0.171          & \cellcolor[HTML]{EFEFEF}0.549          & \cellcolor[HTML]{EFEFEF}0.412          \\ \cline{2-9} 
\multirow{-3}{*}{\textbf{DBN-JIT}}   & \cellcolor[HTML]{CBCEFB}F1           & \cellcolor[HTML]{CBCEFB}0.343          & \cellcolor[HTML]{CBCEFB}0.432          & \cellcolor[HTML]{CBCEFB}0.533          & \cellcolor[HTML]{CBCEFB}0.533          & \cellcolor[HTML]{CBCEFB}0.197          & \cellcolor[HTML]{CBCEFB}0.548          & \cellcolor[HTML]{CBCEFB}0.433          \\ \hline
                                     & ROC                                  & 0.744                                  & 0.749                                  & 0.677                                  & 0.747                                  & 0.750                                  & 0.683                                  & 0.725                                  \\ \cline{2-9} 
                                     & \cellcolor[HTML]{EFEFEF}PR           & \cellcolor[HTML]{EFEFEF}0.319          & \cellcolor[HTML]{EFEFEF}0.413          & \cellcolor[HTML]{EFEFEF}0.644          & \cellcolor[HTML]{EFEFEF}0.533          & \cellcolor[HTML]{EFEFEF}0.184          & \cellcolor[HTML]{EFEFEF}0.549          & \cellcolor[HTML]{EFEFEF}0.440          \\ \cline{2-9} 
\multirow{-3}{*}{\textbf{LApredict}} & \cellcolor[HTML]{CBCEFB}F1           & \cellcolor[HTML]{CBCEFB}0.011          & \cellcolor[HTML]{CBCEFB}0.132          & \cellcolor[HTML]{CBCEFB}0.425          & \cellcolor[HTML]{CBCEFB}0.148          & \cellcolor[HTML]{CBCEFB}0.093          & \cellcolor[HTML]{CBCEFB}0.051          & \cellcolor[HTML]{CBCEFB}0.143          \\ \hline
                                     & ROC                                  & 0.694                                  & 0.713                                  & 0.670                                  & 0.771                                  & 0.703                                  & 0.689                                  & 0.707                                  \\ \cline{2-9} 
                                     & \cellcolor[HTML]{EFEFEF}PR           & \cellcolor[HTML]{EFEFEF}0.280          & \cellcolor[HTML]{EFEFEF}0.396          & \cellcolor[HTML]{EFEFEF}0.644          & \cellcolor[HTML]{EFEFEF}0.574          & \cellcolor[HTML]{EFEFEF}0.135          & \cellcolor[HTML]{EFEFEF}0.569          & \cellcolor[HTML]{EFEFEF}0.433          \\ \cline{2-9} 
\multirow{-3}{*}{\textbf{DeepJIT}}   & \cellcolor[HTML]{CBCEFB}F1           & \cellcolor[HTML]{CBCEFB}0.332          & \cellcolor[HTML]{CBCEFB}0.433          & \cellcolor[HTML]{CBCEFB}0.623          & \cellcolor[HTML]{CBCEFB}0.518          & \cellcolor[HTML]{CBCEFB}0.155          & \cellcolor[HTML]{CBCEFB}0.559          & \cellcolor[HTML]{CBCEFB}0.437          \\ \hline
                                     & ROC                                  & 0.694                                  & 0.723                                  & 0.665                                  & 0.761                                  & 0.699                                  & 0.692                                  & 0.706                                  \\ \cline{2-9} 
                                     & \cellcolor[HTML]{EFEFEF}PR           & \cellcolor[HTML]{EFEFEF}0.279          & \cellcolor[HTML]{EFEFEF}0.394          & \cellcolor[HTML]{EFEFEF}0.640          & \cellcolor[HTML]{EFEFEF}0.559          & \cellcolor[HTML]{EFEFEF}0.134          & \cellcolor[HTML]{EFEFEF}0.577          & \cellcolor[HTML]{EFEFEF}0.431          \\ \cline{2-9} 
\multirow{-3}{*}{\textbf{CC2Vec}}    & \cellcolor[HTML]{CBCEFB}F1           & \cellcolor[HTML]{CBCEFB}0.283          & \cellcolor[HTML]{CBCEFB}0.400          & \cellcolor[HTML]{CBCEFB}0.631          & \cellcolor[HTML]{CBCEFB}0.563          & \cellcolor[HTML]{CBCEFB}0.167          & \cellcolor[HTML]{CBCEFB}0.564          & \cellcolor[HTML]{CBCEFB}0.435          \\ \hline
                                     & ROC                                  & 0.716                                  & 0.747                                  & 0.635                                  & 0.824                                  & 0.764                                  & 0.747                                  & 0.739                                  \\ \cline{2-9} 
                                     & \cellcolor[HTML]{EFEFEF}PR           & \cellcolor[HTML]{EFEFEF}0.292          & \cellcolor[HTML]{EFEFEF}0.408          & \cellcolor[HTML]{EFEFEF}0.614          & \cellcolor[HTML]{EFEFEF}0.659          & \cellcolor[HTML]{EFEFEF}0.168          & \cellcolor[HTML]{EFEFEF}0.646          & \cellcolor[HTML]{EFEFEF}0.465          \\ \cline{2-9} 
\multirow{-3}{*}{\textbf{CodeBERT}}  & \cellcolor[HTML]{CBCEFB}F1           & \cellcolor[HTML]{CBCEFB}0.255          & \cellcolor[HTML]{CBCEFB}0.445          & \cellcolor[HTML]{CBCEFB}0.570          & \cellcolor[HTML]{CBCEFB}0.641          & \cellcolor[HTML]{CBCEFB}0.159          & \cellcolor[HTML]{CBCEFB}0.618          & \cellcolor[HTML]{CBCEFB}0.448          \\ \hline
                                     & ROC                                  & 0.736                                  & 0.748                                  & 0.676                                  & 0.795                                  & 0.725                                  & 0.748                                  & 0.738                                  \\ \cline{2-9} 
                                     & \cellcolor[HTML]{EFEFEF}PR           & \cellcolor[HTML]{EFEFEF}0.317          & \cellcolor[HTML]{EFEFEF}0.428          & \cellcolor[HTML]{EFEFEF}0.648          & \cellcolor[HTML]{EFEFEF}0.589          & \cellcolor[HTML]{EFEFEF}0.153          & \cellcolor[HTML]{EFEFEF}0.650          & \cellcolor[HTML]{EFEFEF}0.464          \\ \cline{2-9} 
\multirow{-3}{*}{\textbf{XGBoost}}   & \cellcolor[HTML]{CBCEFB}F1           & \cellcolor[HTML]{CBCEFB}0.373          & \cellcolor[HTML]{CBCEFB}0.457          & \cellcolor[HTML]{CBCEFB}0.608          & \cellcolor[HTML]{CBCEFB}0.609          & \cellcolor[HTML]{CBCEFB}0.238          & \cellcolor[HTML]{CBCEFB}0.608          & \cellcolor[HTML]{CBCEFB}0.482          \\ \hline
                                     & ROC                                  & 0.739                                  & 0.760                                  & 0.707                                  & 0.798                                  & 0.753                                  & 0.740                                  & 0.750                                  \\ \cline{2-9} 
                                     & \cellcolor[HTML]{EFEFEF}PR           & \cellcolor[HTML]{EFEFEF}0.315          & \cellcolor[HTML]{EFEFEF}0.433          & \cellcolor[HTML]{EFEFEF}0.689          & \cellcolor[HTML]{EFEFEF}0.605          & \cellcolor[HTML]{EFEFEF}0.174          & \cellcolor[HTML]{EFEFEF}0.635          & \cellcolor[HTML]{EFEFEF}0.475          \\ \cline{2-9} 
\multirow{-3}{*}{\textbf{Sim}}       & \cellcolor[HTML]{CBCEFB}F1           & \cellcolor[HTML]{CBCEFB}0.384          & \cellcolor[HTML]{CBCEFB}0.478          & \cellcolor[HTML]{CBCEFB}0.656          & \cellcolor[HTML]{CBCEFB}0.626          & \cellcolor[HTML]{CBCEFB}0.255          & \cellcolor[HTML]{CBCEFB}0.606          & \cellcolor[HTML]{CBCEFB}0.501          \\ \hline
                                     & ROC                                  & 0.757                                  & 0.764                                  & 0.696                                  & 0.814                                  & 0.720                                  & 0.764                                  & 0.753                                  \\ \cline{2-9} 
                                     & \cellcolor[HTML]{EFEFEF}PR           & \cellcolor[HTML]{EFEFEF}0.349          & \cellcolor[HTML]{EFEFEF}0.456          & \cellcolor[HTML]{EFEFEF}0.677          & \cellcolor[HTML]{EFEFEF}0.641          & \cellcolor[HTML]{EFEFEF}0.146          & \cellcolor[HTML]{EFEFEF}0.652          & \cellcolor[HTML]{EFEFEF}0.487          \\ \cline{2-9} 
\multirow{-3}{*}{\textbf{Com}}       & \cellcolor[HTML]{CBCEFB}F1           & \cellcolor[HTML]{CBCEFB}0.394          & \cellcolor[HTML]{CBCEFB}0.457          & \cellcolor[HTML]{CBCEFB}0.619          & \cellcolor[HTML]{CBCEFB}0.633          & \cellcolor[HTML]{CBCEFB}0.222          & \cellcolor[HTML]{CBCEFB}0.634          & \cellcolor[HTML]{CBCEFB}0.493          \\ \hline
\end{tabular}
\label{rq1}
}
\end{table}

To answer this RQ, we carry out an experiment by adopting typical JIT defect prediction baselines as well as our component models to show the effectiveness of expert knowledge-based and deep learning-based approaches.

Table~\ref{rq1} shows the performance of all the compared approaches in terms of AUC-ROC, AUC-PR, and F1-scores respectively.
In Table~\ref{rq1}, we categorize approaches into five groups: 
\begin{itemize}
    \item \textbf{Expert knowledge-based baselines}: LR-JIT, DBN-JIT, and LApredict;

    \item \textbf{Deep learning-based baselines}: DeepJIT, CC2Vec, and CodeBERT

    \item \textbf{A popular ensemble learning baseline}: Extreme Gradient Boosting (XGBoost)~\citep{chen2016xgboost}

     \item \textbf{Our component models}: Sim and Com

\end{itemize}

The experimental results demonstrate that our component models (i.e., \emph{Sim} and \emph{Com}) outperform the expert knowledge-based baselines and deep learning-based baselines. Specifically, \emph{Sim} outperforms other expert knowledge-based baselines by \revised{1.6}\%--\revised{9.6}\%, \revised{2.3}\%-\revised{15.9}\%, and \revised{3.9}\%--\revised{250.3}\% in terms of the AUC-ROC, AUC-PR, and F1 score on average.
\emph{Com} outperforms other deep learning-based baselines by 1.9\%--6.7\%,4.7\%-13.0\%, and 10.0\%--13.3\% in terms of the AUC-ROC, AUC-PR, and F1 score on average.
The results indicate that \emph{Sim} and \emph{Com} are the two best models in expert knowledge-based baselines and deep learning-based approaches respectively (excluding models after fusion).

\find{\textbf{Finding 1:} If without model fusion techniques, \emph{Sim} and \emph{Com} are the two best-performing models for deep learning-based and expert knowledge-based JIT approaches respectively.}

Moreover, we observe that the pre-trained model of code (i.e., CodeBERT) outperforms the prior deep learning-based baselines (i.e., DeepJIT and CC2Vec) for most cases.
However, CodeBERT fails to outperform \emph{Com} in most cases: it only outperforms \emph{Com} in Platform and Gerrit datasets.
On average, \emph{Com} outperforms CodeBERT by 1.9\%, 4.7\%, and 10.0\% for AUC-ROC, AUC-PR, and F1 scores respectively.
This indicates that the pre-trained model of code is not able to always defeat deep learning models without pre-training, especially when the deep learning model has tailored designs (e.g., hierarchical ``line$\rightarrow$hunk$\rightarrow$file$\rightarrow$commit'' feature extractions) for the task.

\revised{To present the effectiveness of the hierarchical feature extraction, we trained two variants of \emph{Com}: one using this hierarchical feature extraction, and the other without it. Table~\ref{ablation_com2} displays the average performance metrics on test sets, including AUC-ROC, AUC-PR, and F1 scores. Results showed that the hierarchical feature extraction can lead to 2.0\%, 3.4\%, and 1.9\% improvements in AUC-ROC, AUC-PR, and F1 respectively.}

\begin{table}[h]
\caption{The performance of Com when removing the hierarchical feature extraction}
\centering
\begin{tabular}{|l|c|c|c|}
\hline
\textbf{Averaged Results on Test Set}       & \textbf{AUC-ROC} & \textbf{AUC-PR} & \textbf{F1}   \\ \hline
\textbf{Com}                                 & \textbf{75.3}    & \textbf{48.7}   & \textbf{49.3} \\ \hline
\textbf{Com w/o  hierarchical feature} & 73.8             & 47.1            & 48.4          \\ \hline
\end{tabular}
\label{ablation_com2}
\end{table}

\find{\textbf{Finding 2:} The pre-trained model of code cannot lead to improvements over \emph{Com} on average. \revised{Besides, the hierarchical feature extraction of \emph{Com} helps to extract more effective features.}}

In addition, we observe that the superiority of \emph{Sim} and \emph{Com} varies with different projects and different evaluation metrics. For instance, \emph{Sim} performs better for OpenStack (in F1 scores), JDT (in AUC-ROC, AUC-PR, and F1 scores), and Gerrit (AUC-ROC, AUC-PR, and F1 scores) while \emph{Com} performs better in other projects and metrics.

\find{\textbf{Finding 3:} The superiority of Sim and Com varies with different projects and different evaluation metrics which may imply that the combination of the two could further improve the effectiveness.}

\revised{Furthermore, DeepJIT and Com share a lot of similarities. For example, they both use textCNN as the feature extractor and use the same hierarchical feature extraction from code lines to commits. However, Com substantially outperforms DeepJIT. Here, we highlight two main differences between DeepJIT and our Com model in the following:}

\begin{itemize}
    \item \revised{Com highlights the property of code change (either ``added'' or ``deleted'') by adding the special headers while DeepJIT does not distinguish the added or deleted code.}
    \item \revised{Com selects the best model checkpoint for testing based on the performance of the validation set (in terms of the average of three evaluation metrics). In contrast, DeepJIT selects the best model checkpoint based on the lowest loss value. Our model selection strategy is more robust and effective than DeeJIT’s model selection strategy.}
\end{itemize}

\revised{To present the effectiveness of these two improvements proposed by us, we add the additional ablation studies, and the results are shown in Table~\ref{ablation_com}. Results showed that the two improvements enhanced model performance. If we remove our proposed two improvements, the performance of Com will drop 6.5\%, 12.5\%, and 12.8\% in AUC-ROC, AUC-PR, and F1 respectively.}

\begin{table}[t]
\caption{Ablation studies of Com on the test set}
\centering
\begin{tabular}{|l|c|c|c|}
\hline
\textbf{Averaged Results on Test Set}       & \textbf{AUC-ROC} & \textbf{AUC-PR} & \textbf{F1}   \\ \hline
\textbf{Com}                                 & \textbf{75.3}    & \textbf{48.7}   & \textbf{49.3} \\ \hline
\textbf{\hspace{0.2em} -w/o headers}                           & 74.8             & 47.8            & 49.2          \\ \hline
\textbf{\hspace{0.2em} -w/o headers \& our model selection}         & 70.7             & 43.3            & 43.7          \\ \hline
\end{tabular}
\label{ablation_com}
\end{table}

\find{\textbf{Finding 4:} The superiority of Com over DeepJIT comes from 1) the better pre-processing to highlight code changes and 2) the better model snapshot selection strategy.}

\begin{center}
\begin{tcolorbox}[colback=gray!10,
                  colframe=black,
                  width=8.5cm,
                  arc=1mm, auto outer arc,
                  boxrule=0.5pt,
                 ]
\textbf{Answers to RQ1:}  Without using model fusion techniques, \emph{Sim} and \emph{Com} are the best-performing models for JIT defect prediction. The superiority of \emph{Sim} and \emph{Com} varies, indicating that combining these two models may obtain a better solution.  
\end{tcolorbox}
\end{center}

\subsection{\textbf{RQ2: How different are the predictions of expert knowledge-based and deep learning-based approaches?}}

In this RQ, we aim to explore the difference between the predictions of expert knowledge-based tools and deep learning-based tools. 
If their predictions are different from each other, it indicates their correlation is relatively low and it is more likely to get improvements if fusing them.
Specifically, we focus on analyzing the \emph{Sim} and \emph{Com} because they are not only the best-performing models in their own categories (identified in RQ1) but also the building components for our model fusion model. 
In Table~\ref{rq1}, we find that the superiority of \emph{Sim} and \emph{Com} varies with different projects and different evaluation metrics. 
However, this high-level observation is not enough to show how different these two models' behaviors are.

\begin{figure}
\begin{minipage}[t]{0.5\textwidth}
    \centering
    \includegraphics[scale=0.33]{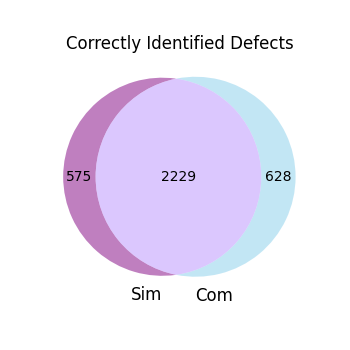}
  \end{minipage}%
  \begin{minipage}[t]{0.5\textwidth}
    \centering
    \includegraphics[scale=0.33]{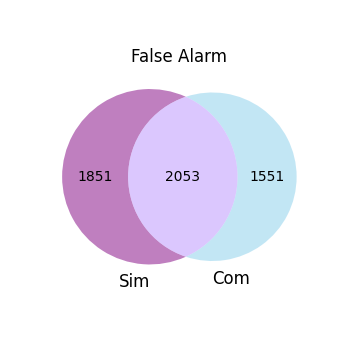}
  \end{minipage}%
  \vspace{-0.2cm}
  \caption{Relations between the predictions of Sim and Com. The left Venn diagram shows the identified defective commits and the right one shows the false positives generated by Sim and Com.}
  \label{rq2_venn}
\end{figure}

First, we visualize the prediction results of \emph{Sim} and \emph{Com} via Venn diagrams.
As shown in Figure~\ref{rq2_venn}, although \emph{Sim} and \emph{Com} identify 2,229 common defective commits, they also identify 575 (20.5\%) and 628 (22.0\%) different defective commits respectively.  
In addition, \emph{Sim} and \emph{Com} produce 1,851 and 1,551 different false positives, accounting for 47.4\% and 43.0\% of their false positives respectively. 
The results above show that the predictions of \emph{Sim} and \emph{Com} are different from each other.
For instance, 20.5\% of defective commits identified by \emph{Sim} are not identified by \emph{Com}, and 47.4\% of false positives are only produced by \emph{Sim} but not \emph{Com}.

\find{\textbf{Finding 5:} Sim and Com identify 20.5\% and 22.0\% unique defective commits and produce 47.4\% and 43.0\% unique false positives, demonstrating their predictions are quite different.}

\begin{figure}[thb] 
\centering 
\includegraphics[width=0.8\textwidth]{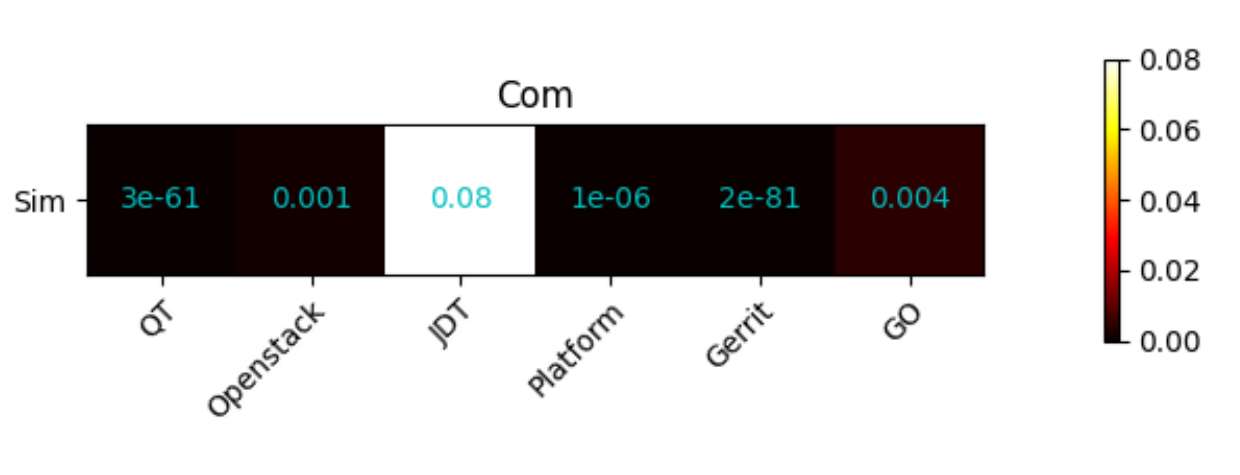}
\caption{P-values between the expert knowledge-based approach (Sim) and the deep learning-based approach (Com).}
\label{rq2_p_value} 
\end{figure}

Second, we carry out Wilcoxon signed-rank tests with the null hypothesis that predictions generated by \emph{Sim} and \emph{Com} follow the same distribution. 
As shown in Figure~\ref{rq2_p_value}, the p-values between \emph{Sim} and \emph{Com} are larger than 0.05 only for the JDT dataset, indicating for most datasets, the predictions of Sim and Com are significantly different.

\find{\textbf{Finding 6:} For most datasets, predictions of Sim and Com are significantly different in statistics.}

\begin{center}
\begin{tcolorbox}[colback=gray!10,
                  colframe=black,
                  width=8.5cm,
                  arc=1mm, auto outer arc,
                  boxrule=0.5pt,
                 ]
\textbf{Answers to RQ2:}  We confirmed that the predictions of \emph{Sim} and \emph{Com} are quite different from each other: 1) they both identify many unique defects and produce many unique false positives, and 2) their predictions are significantly different for most datasets.
\end{tcolorbox}
\end{center}
\subsection{\textbf{RQ3:  What are the best model fusion strategies in the JIT
defect prediction?}}

In this RQ, we aim to explore and identify the best model fusion combination and experiment with our final model.
\revised{Firstly, we will identify the best simple and complex models to fully leverage the hand-crafted features and commit contents, respectively. Secondly, we iterate on all possible combinations on the chosen simple and complex models to form 20 model variants and do validation on each model variant. By doing so, we can identify the best model fusion combination based on the performances of the validation sets. Lastly, we adopt the found best model fusion strategies to form our SimCom++ model,  evaluate its effectiveness in the test set,  and compare with existing JIT defect prediction baselines.}

\subsubsection{\revised{Identifying Component Models}}

\revised{For model fusion, in most cases, the stronger the constituent models, the more likely it is to achieve the best results after fusion. Therefore, we aim to identify the best traditional Machine Learning classifier and Deep Learning model as the simple and complex component models respectively.}

\revised{For the simple model dealing with the hand-crafted features, we consider the following widely used traditional Machine Learning classifiers: Logistic Regression (LR), Decision Tree (DT), Support Vector Machine (SVM), Naive Bayes (NB), XGBoost, Random Forest (RF). Specifically, we trained all candidate machine learning classifiers on the training sets and assessed their performance on the validation sets. Table~\ref{rq3_valid_simple} presents the average performances of each candidate model in terms of AUC-ROC, AUC-PR, and F1. We found that Random Forest consistently outperformed other ML classifiers in three metrics. Thus, we have decided to utilize Random Forest as the simple model for its better performance in validation sets. One plausible explanation for RF's better performance is its ensemble structure, consisting of multiple ML models (i.e., decision trees). This ensemble characteristic contributes to reducing overfitting and enhancing the model's ability to generalize effectively~\citep{zhang2012ensemble}.}

\begin{table}[h]
 \centering
\caption{The averaged performance of multiple ML classifiers in the validation sets}
\resizebox{\columnwidth}{!}{%
\begin{tabular}{|l|c|c|c|c|c|c|}
\hline
    \textbf{Validation} & \multicolumn{1}{l|}{\textbf{Logistic Regression}} & \multicolumn{1}{l|}{\textbf{Decision Tree}} & \multicolumn{1}{l|}{\textbf{SVM}} & \multicolumn{1}{l|}{\textbf{Naïve Bayes}} & \multicolumn{1}{l|}{\textbf{XGBoost}} & \multicolumn{1}{l|}{\textbf{Random Forest}} \\ \hline
\cellcolor[HTML]{FFFFFF} ROC                 & \cellcolor[HTML]{FFFFFF} 67.2                                              &\cellcolor[HTML]{FFFFFF} 71.3                                        & \cellcolor[HTML]{FFFFFF} 58.9                                                 & \cellcolor[HTML]{FFFFFF} 68.3                                      & \cellcolor[HTML]{FFFFFF} 73.6                                  & \cellcolor[HTML]{FFFFFF} \textbf{76.0}                                        \\ \hline
 \cellcolor[HTML]{EFEFEF} PR                  &  \cellcolor[HTML]{EFEFEF} 43.8                                              &  \cellcolor[HTML]{EFEFEF} 45.2                                       &  \cellcolor[HTML]{EFEFEF} 50.1                                                 &  \cellcolor[HTML]{EFEFEF} 43.5                                      &  \cellcolor[HTML]{EFEFEF} 48.7                                  &  \cellcolor[HTML]{EFEFEF} \textbf{50.4}                                        \\ \hline
\cellcolor[HTML]{CBCEFB} F1                  & \cellcolor[HTML]{CBCEFB} 45.0                                              & \cellcolor[HTML]{CBCEFB} 50.1                                        & \cellcolor[HTML]{CBCEFB} 41.5                                                 & \cellcolor[HTML]{CBCEFB} 38.0                                      & \cellcolor[HTML]{CBCEFB} 50.1                                  & \cellcolor[HTML]{CBCEFB} \textbf{51.1}                                        \\ \hline
\end{tabular}
\label{rq3_valid_simple}
}
\end{table}

\revised{
For the complex model Com, we explore the use of two popular Deep Learning models textCNN and Transformers, as potential feature extractors. We aim to select one feature extractor that demonstrates better performance on the validation sets. To achieve this, we separately trained two Com variants: one utilizing textCNN and the other employing Transformers. Subsequently, we assessed the performance of these two variants on the validation sets. Table~\ref{rq3_valid_complex} displays the average performance metrics on validation sets, including AUC-ROC, AUC-PR, and F1 scores, for each candidate model. The results show that  textCNN consistently outperformed Transformers in all metrics. Therefore, we opted to employ textCNN as the feature extractor for the complex model, which is aligned with the model descriptions in Section 3.2.}

\begin{table}[h]
\centering
\caption{The averaged performance of different DL feature extractors in the validation sets}
\resizebox{0.5\columnwidth}{!}{%
\begin{tabular}{|l|c|c|}
\hline
\textbf{Validation sets} & \multicolumn{1}{l|}{\textbf{Transformer}} & \multicolumn{1}{l|}{\textbf{textCNN}} \\ \hline
\cellcolor[HTML]{FFFFFF} ROC                      & \cellcolor[HTML]{FFFFFF} 75.0                                      & \cellcolor[HTML]{FFFFFF} \textbf{78.0}                         \\ \hline
\cellcolor[HTML]{EFEFEF} PR                       & \cellcolor[HTML]{EFEFEF} 50.7                                      & \cellcolor[HTML]{EFEFEF} \textbf{54.7}                         \\ \hline
\cellcolor[HTML]{CBCEFB} F1                       & \cellcolor[HTML]{CBCEFB} 51.5                                      & \cellcolor[HTML]{CBCEFB} \textbf{53.9}                         \\ \hline
\end{tabular}
\label{rq3_valid_complex}
}
\end{table}

\subsubsection{\revised{Identifying Best Model Fusion}}
\begin{figure}[h] 
\centering 
\includegraphics[width=0.8\textwidth]{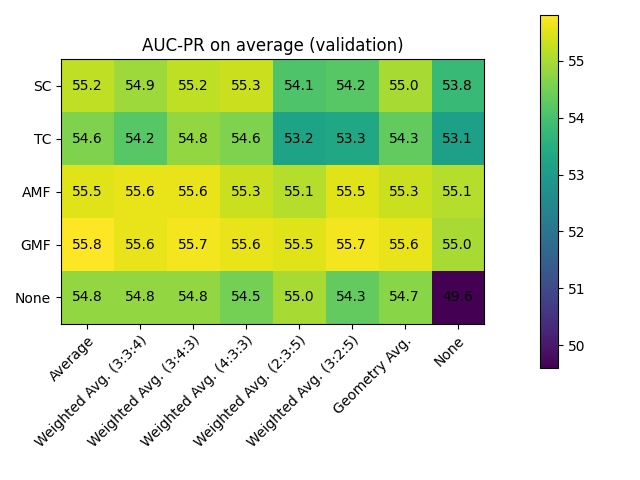}
\caption{Validation results of different early/late fusion combinations in AUC-PR. The x-axis is the late fusion and the y-axis refers to the early fusion.}
\label{rq3_valid_result} 
\end{figure}

Here, we combine the component models Sim and Com via model fusion strategies. By adopting one model fusion strategy, we will get a model variant. 
We regard the model variant that has the best validation performance as our final model namely SimCom++.
We call it SimCom++ because it combines Sim and Com and adopts two model fusion strategies (early fusion and late fusion) to augment it.
Figure~\ref{fusion_combination} presents the overall framework of our proposed approach. The best model can be determined by specifying the following choices:
\begin{itemize}
    \item{Early fusion strategy (SC, TC, AMF, GMF, and None); }
    \item{Late fusion strategy (Average, Weighted Average, Geometric Average, and None). }
\end{itemize}
We choose the combination that leads to the best performance \revised{(in terms of AUC-PR)} in validation to form the final proposed model. \revised{We chose AUC-PR as the target metric because it is more suitable for the imbalanced dataset~\citep{Davis2006TheRB}.
}
As shown in Figure~\ref{rq3_valid_result}, we can obtain the best validation result when adopting Gating Mechanism-based Fusion (GMF) as the early fusion and the average strategy as the late fusion. 
\revised{Thus, we identify the ``GMF+average'' combination as the best model fusion strategy.}

\revised{Besides, as shown in Figure 7, we found that AMF and GMF behave similarly in the validation set (with an average difference of 0.2 AUC-PR score). But we can still obervse some differences between them. For example, GMF performs best with the combination of “Average” while AMF performs best with the combination of “Weighted Average”. This suggests that AMF and GMF do not behave similarly in all aspects.}

\find{\textbf{Finding 7:} Experimental results show that Gating Mechanism-based Fusion is the best early fusion strategy and the average is the best late fusion strategy.}

\begin{table}[]
 \centering
\caption{The performance of SimCom++ and its component models with respect to AUC-ROC, AUC-PR, and F1 scores}
\resizebox{\columnwidth}{!}{%
\begin{tabular}{|l|l|r|r|r|r|r|r|r|}
\hline
\textbf{Tools}                    & \multicolumn{1}{c|}{\textbf{Metric}} & \multicolumn{1}{c|}{\textbf{QT}}       & \multicolumn{1}{c|}{\textbf{OP.}}      & \multicolumn{1}{c|}{\textbf{JDT}}      & \multicolumn{1}{c|}{\textbf{Plat.}}    & \multicolumn{1}{c|}{\textbf{Gerrit}}   & \multicolumn{1}{c|}{\textbf{GO}}       & \multicolumn{1}{c|}{\textbf{Avg.}}     \\ \hline
                                     & ROC                                  & 0.739                                  & 0.760                                  & 0.707                                  & 0.798                                  & 0.753                                  & 0.740                                  & 0.750                                  \\ \cline{2-9} 
                                     & \cellcolor[HTML]{EFEFEF}PR           & \cellcolor[HTML]{EFEFEF}0.315          & \cellcolor[HTML]{EFEFEF}0.433          & \cellcolor[HTML]{EFEFEF}0.689          & \cellcolor[HTML]{EFEFEF}0.605          & \cellcolor[HTML]{EFEFEF}0.174          & \cellcolor[HTML]{EFEFEF}0.635          & \cellcolor[HTML]{EFEFEF}0.475          \\ \cline{2-9} 
\multirow{-3}{*}{\textbf{Sim}}       & \cellcolor[HTML]{CBCEFB}F1           & \cellcolor[HTML]{CBCEFB}0.384          & \cellcolor[HTML]{CBCEFB}0.478          & \cellcolor[HTML]{CBCEFB}0.656          & \cellcolor[HTML]{CBCEFB}0.626          & \cellcolor[HTML]{CBCEFB}0.255          & \cellcolor[HTML]{CBCEFB}0.606          & \cellcolor[HTML]{CBCEFB}0.501          \\ \hline
                                     & ROC                                  & 0.757                                  & 0.764                                  & 0.696                                  & 0.814                                  & 0.720                                  & 0.764                                  & 0.753                                  \\ \cline{2-9} 
                                     & \cellcolor[HTML]{EFEFEF}PR           & \cellcolor[HTML]{EFEFEF}0.349          & \cellcolor[HTML]{EFEFEF}0.456          & \cellcolor[HTML]{EFEFEF}0.677          & \cellcolor[HTML]{EFEFEF}0.641          & \cellcolor[HTML]{EFEFEF}0.146          & \cellcolor[HTML]{EFEFEF}0.652          & \cellcolor[HTML]{EFEFEF}0.487          \\ \cline{2-9} 
\multirow{-3}{*}{\textbf{Com}}       & \cellcolor[HTML]{CBCEFB}F1           & \cellcolor[HTML]{CBCEFB}0.394          & \cellcolor[HTML]{CBCEFB}0.457          & \cellcolor[HTML]{CBCEFB}0.619          & \cellcolor[HTML]{CBCEFB}0.633          & \cellcolor[HTML]{CBCEFB}0.222          & \cellcolor[HTML]{CBCEFB}0.634          & \cellcolor[HTML]{CBCEFB}0.493          \\ \hline
                                     & ROC                                  & \textbf{0.777}                         & \textbf{0.786}                         & \textbf{0.730}                         & \textbf{0.830}                         & \textbf{0.773}                         & \textbf{0.790}                         & \textbf{0.781}                         \\ \cline{2-9} 
                                     & \cellcolor[HTML]{EFEFEF}PR           & \cellcolor[HTML]{EFEFEF}\textbf{0.390} & \cellcolor[HTML]{EFEFEF}\textbf{0.470} & \cellcolor[HTML]{EFEFEF}\textbf{0.708} & \cellcolor[HTML]{EFEFEF}\textbf{0.675} & \cellcolor[HTML]{EFEFEF}\textbf{0.195} & \cellcolor[HTML]{EFEFEF}\textbf{0.698} & \cellcolor[HTML]{EFEFEF}\textbf{0.523} \\ \cline{2-9} 
\multirow{-3}{*}{\textbf{SimCom++}}  & \cellcolor[HTML]{CBCEFB}F1           & \cellcolor[HTML]{CBCEFB}\textbf{0.419} & \cellcolor[HTML]{CBCEFB}\textbf{0.492} & \cellcolor[HTML]{CBCEFB}\textbf{0.659} & \cellcolor[HTML]{CBCEFB}\textbf{0.659} & \cellcolor[HTML]{CBCEFB}\textbf{0.273} & \cellcolor[HTML]{CBCEFB}\textbf{0.664} & \cellcolor[HTML]{CBCEFB}\textbf{0.528} \\ \hline
\end{tabular}
\label{rq3}
}
\end{table}

\subsubsection{\revised{SimCom++ V.S. Existing Methods}}

\begin{table}[]
\caption{Ablation studies of SimCom++ on the test set}
\begin{tabular}{|l|c|c|c|}
\hline
\textbf{Averaged Results on Test Set}              & \textbf{AUC-ROC} & \textbf{AUC-PR} & \textbf{F1}   \\ \hline
\textbf{SimCom++}               & \textbf{78.1}    & \textbf{52.3}   & \textbf{52.8} \\ \hline
-w/o late fusion         & 77.4             & 51.5            & 51.9          \\ \hline
-w/o early fusion         & 77.1             & 50.9            & 50.8          \\ \hline
-w/o late \& early fusions & 75.0             & 47.5            & 50.1          \\ \hline
\end{tabular}
\label{ablation_simcom_plus}
\end{table}

As shown in Table~\ref{rq3}, by fusing the expert knowledge-based and deep learning-based models, the combined model SimCom++ leads to consistently better (4.1\%, 10.1\%, 7.1\% on average and up to 6.8\%, 33.6\%,23.0\% in AUC-ROC, AUC-PR, and F1-scores respectively) performance than its component models.

SimCom++ also outperforms all other baselines from 5.7\% to 26.9\% in different metrics.
For instance, SimCom++ leads to improvements over the existing best expert knowledge-based approach LAPredict by 7.7\% and 18.9\% in AUC-ROC and AUC-PR.
SimCom++ also outperforms the latest deep learning-based approach CC2Vec by 10.6\% and 21.3\% in AUC-ROC and AUC-PR. The results show the effectiveness of SimCom++ over other existing approaches.
\revised{We conduct the Wilcoxon Signed Rank at a 95\% significance level (i.e., p-value $<$ 0.05) and calculate Cliff’s delta on the paired data corresponding to SimCom++ and the best-performing baseline model in each evaluation metric (i.e., CodeBERT for AUC-ROC, CodeBERT for AUC-PR, and XGBoost in F1). 
Our rationale is that if SimCom++ cannot significantly outperform any baseline, the best performer in each metric has the highest chance to be the baseline.
The significance test has been conducted on the values of evaluation metrics. Specifically, when assessing F1 scores, we compute the F1 score for each sample within the test set and analyze the distribution of these scores. However, when dealing with AUC-ROC and AUC-PR, which measure the area under curves, it's crucial to have at least one sample from one class to derive a valid metric value. Consequently, it is not possible to calculate AUC-PR and AUC-ROC metrics for a single sample. To address this limitation, we randomly split the test set into 10 groups. Within each group, we calculate the AUC-ROC and AUC-PR metrics based on the available samples. This approach ensures that valid AUC-ROC and AUC-PR values are obtained. 
For Cliff’s delta, we consider a delta that is less than 0.147, between 0.147 and 0.33, between 0.33 and 0.474, and above 0.474 as Negligible (N), Small (S), Medium (M), and Large (L) effect size, respectively following previous literature~\citep{cliff2014ordinal}. We observed that p-values are less than 0.05 and Cliff’s delta are from 0.33 to 0.56, which correspond to the Medium or Large sizes. These results indicate that SimCom++ significantly and substantially outperforms the best-performing baseline models. 
Furthermore, a recent study on when to use the Bonferroni correction~\citep{armstrong2014use} advised that if the study is restricted to a small number of planned comparisons, no correction would be advised to be conducted. In our experiments, we only compared SimCom++ with the best-performing baselines, which is a relatively small number of comparisons and does not need the Bonferroni correction.}

\revised{SimCom++ outperformed the baselines because of two reasons: 1) the model fusion strategies and 2) the improved simple and complex models. 
As shown in Table~\ref{ablation_simcom_plus},  our model fusion has improved the component models by 4.0\%, 10.1\%, and 5.4\% in AUC-ROC, AUC-PR, and F1, respectively. This highlights the contribution of our model fusion in effectiveness. Besides, our component models Sim and Com have improved substantially on the baselines, which also make the contribution to the effectiveness of the final fused model. As shown in Table~\ref{rq1}, Sim improves the best simple model baseline XGBoost by 1.6\%, 2.3\%, and 3.9\% in AUC-ROC, AUC-PR, and F1. Com improves DeepJIT (a complex model baseline similar to Com) by 6.5\%, 12.5\%, and 12.8\% in AUC-ROC, AUC-PR, and F1.
}

\begin{center}
\begin{tcolorbox}[colback=gray!10,
                  colframe=black,
                  width=8.5cm,
                  arc=1mm, auto outer arc,
                  boxrule=0.5pt,
                 ]
\textbf{Answers to RQ3:}
We identify ``GMF+average'' as the best model fusion strategy and our final model SimCom++ performs consistently better (about 4.1\%--10.1\%) than its component models and outperforms the baselines by 5.7\%--26.9\%.
\end{tcolorbox}
\end{center}

\subsection{\textbf{RQ4:  \revised{Why does the model fusion lead to better effectiveness?}}}

This RQ aims to explain how the model fusion works in the JIT defect prediction. 
Specifically, we focus on analyzing the model behavior differences between component models (\emph{Sim} and \emph{Com}) and the final fused model SimCom++.

\begin{figure}
\begin{minipage}[t]{0.5\textwidth}
    \centering
    \includegraphics[scale=0.33]{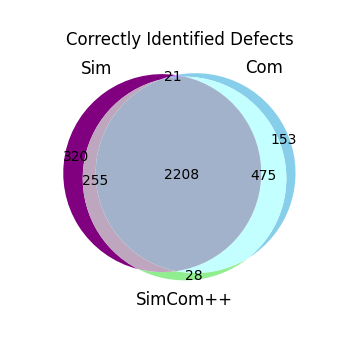}
  \end{minipage}%
  \begin{minipage}[t]{0.5\textwidth}
    \centering
    \includegraphics[scale=0.33]{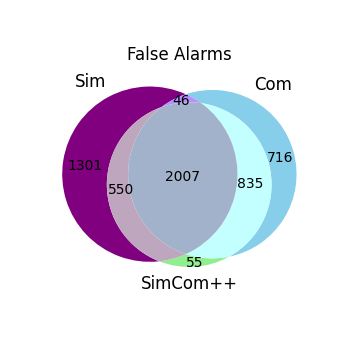}
  \end{minipage}%
  \vspace{-0.2cm}
  \caption{Relations between the predictions of Sim, Com, and SimCom++. The left Venn diagram shows the identified defective commits and the right one shows the false positives.}
  \label{venn_simcom}
\end{figure}

First, we visualize the prediction results of Sim, Com, and SimCom++ via Venn diagrams. Figure~\ref{venn_simcom} shows that SimCom++ cannot identify more unique defective commits that are not identified by Sim and Com. \revised{However, SimCom++ can merge the correct predictions of Sim and Com and recognize even more defective commits (2,966) than Sim (2,804) and Com (2,857). In addition, SimCom++ produces relatively few new false positives (only 55 commits) with respect to Sim and Com. SimCom++ can reduce the false positives from 3,904 for Sim and 3,604 for Com to 3,447, resulting in 11.7\% and 4.3\% fewer false positives.}

\find {\textbf{Finding 8:} Model fusion strategies could help to reduce 11.7\% to 4.3\% of the false positives of Sim and Com respectively. }

Second, we analyze to what extent SimCom++ can correct the wrong predictions of its component models. To obtain the prediction class, we turn prediction scores into classes by the following commonly used rule~\citep{Kamei2013ALE}: if a prediction score is larger than 0.5, the corresponding prediction class is ``defective''. Otherwise, the prediction class is ``clean''.
For ease of analysis, we merge six project datasets as the whole data. 
\revised{Firstly, by checking the predictions made by Sim and Com respectively, we found that out of a total of 18,986 predictions across all test sets, Sim and Com diverge in 4,406 instances (23.2\%). Within these 4,406 discrepant predictions, Sim correctly predicts 2,020 cases (10.6\%) while making 2,386 incorrect predictions (12.6\%).}
Then we count the number of commits where SimCom++ and its component models have different prediction classes. Furthermore, we check the number of commits where SimCom++ corrects the wrong predictions made by \emph{Sim} or \emph{Com}. Also, we checked whether SimCom++ will distort the original accurate predictions into the wrong ones as presented in Table~\ref{rq4}. 
``\#Wrong$\rightarrow$Correct'' refers to the number of commits where SimCom++ is correct but \emph{Sim}/\emph{Com} is wrong.
Similarly, ``\#Correct$\rightarrow$Wrong'' refers to the number of commits where SimCom++ is wrong but \emph{Sim}/\emph{Com} is correct.
Besides, ``\#Net Correction'' refers to ``\#Wrong$\rightarrow$Correct'' minus ``\#Correct$\rightarrow$Wrong''. ``Net Correction Ratio'' refers to ``\#Net Correction'' divided by ``\#Different Predictions''.
Comparing SimCom++ and Sim, they have different prediction classes for 3,113 and among those cases, SimCom++ can help to correct 601 wrong predictions of \emph{Sim}, accounting for 19.3\% of 3,113 commits.
For \emph{Com}, SimCom++ could correct 234 wrong predictions, accounting for 12.0\% of 1,945 commits.

\find {\textbf{Finding 9:} When predictions are different between SimCom++ and \emph{Sim} (\emph{Com}),  for 19.3\% and 12.0\% of cases, SimCom++ are correcting the wrong predictions made by \emph{Sim} and \emph{Com} via model fusion methods. }

\begin{table}[t]
\centering
\caption{Analysis of different prediction classes between SimCom++ and its component models (Sim and Com).}
\resizebox{0.65\columnwidth}{!}{%
\begin{tabular}{|l|c|c|}
\hline
\textbf{Different Prediction Analysis} & \textbf{Sim} & \multicolumn{1}{l|}{\textbf{Com}} \\ \hline
\#Different Predictions       & 3,113        & 1,945                             \\ \hline
\#Wrong$\rightarrow$Correct & 1,857        & 1,087                             \\ \hline
\#Correct$\rightarrow$Wrong & 1,256        & 853                               \\ \hline
\#Net Correction              & 601          & 234                               \\ \hline
Net Correction Ratio (\%)         & 19.3\%       & 12.0\%                            \\ \hline
\end{tabular}
}
\label{rq4}
\end{table}

\begin{figure}[t] 
\centering 
\includegraphics[width=0.9\textwidth]{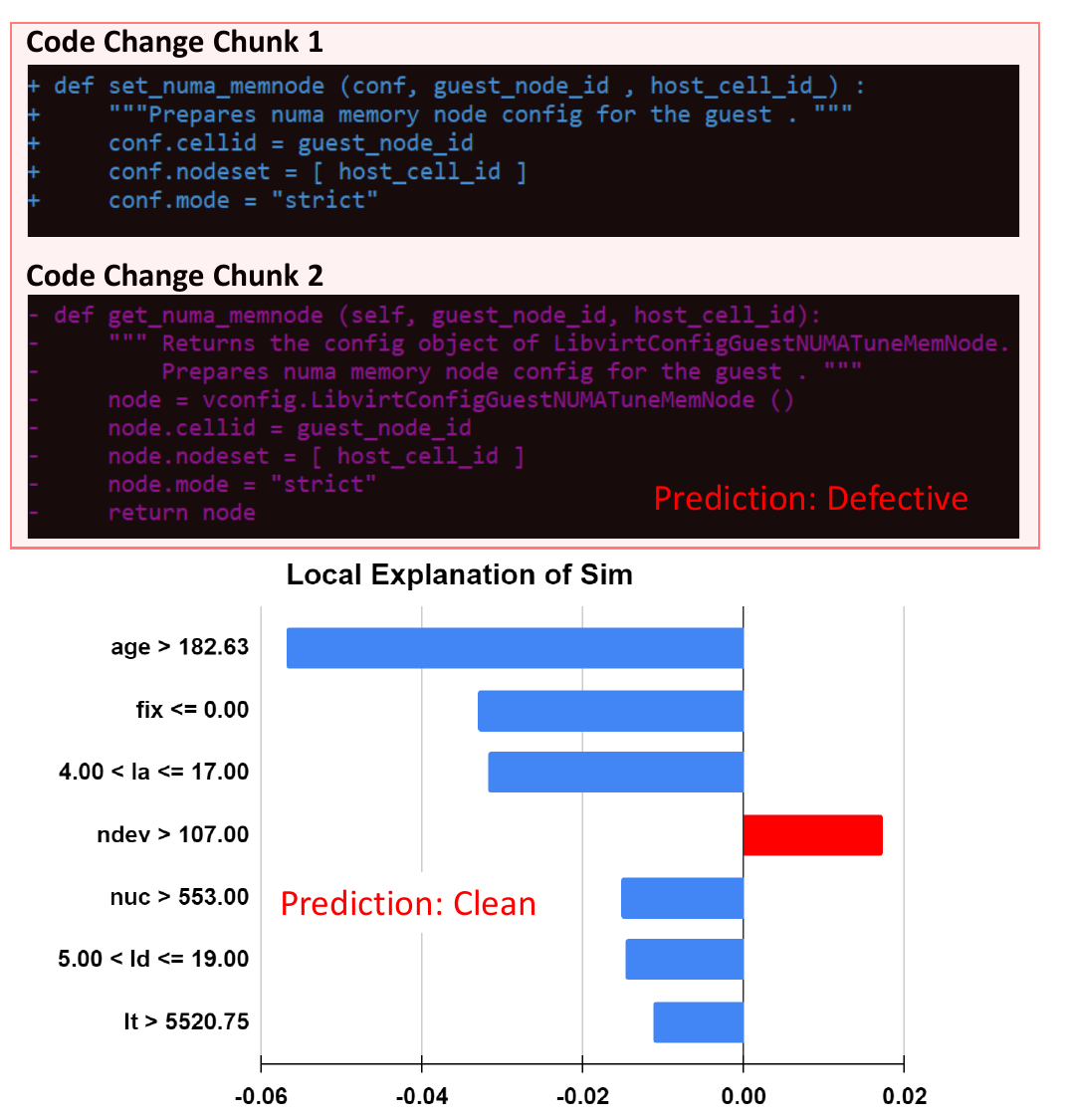}
\caption{A case study on a commit from the {\tt OpenStack} project. The top part is the code changes and Com predicts the commit as ``defective'' while Sim as well as SimCom++ predicts it as ``clean''. The bottom part is the explanations of Sim on its prediction by adopting the local explanation tool LIME~\citep{LIME}.}
\label{rq4_qualtatitive_analysis} 
\end{figure}

Third, we investigate how expert knowledge helps deep learning-based models. 
Figure~\ref{rq4_qualtatitive_analysis} illustrates an example of a commit taken from the {\tt OpenStack} project. The commit (code changes on the top) edits the function by renaming variables and functions and deleting the return statement.
The deep learning-based model \emph{Com} initially predicts the commit as ``defective''. However, this commit is a clean commit in fact.
By fusing expert knowledge into \emph{Com}, the prediction result is changed to ``clean'' which is in line with the ground truth. To understand how expert knowledge affects deep learning models' decisions, we adopt a popular local model-agnostic explanation tool namely LIME~\citep{LIME} on \emph{Sim} to show how expert knowledge leads to the ``clean'' prediction.
\revised{As shown in the bottom part of Figure~\ref{rq4_qualtatitive_analysis}, we provide an explanation of the expert knowledge associated with the top three expert-created features, specifically ``age'', ``fix'', and ``la'' as these features make the most substantial contributions to the prediction:}

\begin{itemize} [leftmargin=*]
\item \revised{\textit{``age" (Age of Developers):}} It considers the module where the code changes happen is old and has not been changed for a long time (``age $>$ 182.63''), which may have fewer flaws according to~\citep{Graves2000PredictingFI}. 

\item \revised{\textit{``fix" (Is Bug-fixing Commit or Not):}} It indicates whether a commit aims to fix a bug or not. As reported in~\citep{Purushothaman2005TowardUT}, nearly 40\% of bug-fixing commits introduced one or more other defects. However, this commit is not a bug-fixing commit (``fix $<=$ 0.00''). 

\item \revised{\textit{``la" (Added Code Lines):}} It denotes the quantity of added code lines. This commit does not add many new code lines (``4.00 $<$ la $<=$ 17.00''). The less new code added, the lower the possibility of making mistakes, c.f.,~\citep{Moser2008ACA, Nagappan2005UseOR}.

\end{itemize}

\revised{In Figure~\ref{rq4_qualtatitive_analysis}, it's evident that the remaining expert-crafted features contribute less in comparison to the aforementioned top three features. Here are the elaborations on these features:}

\begin{itemize} [leftmargin=*]

\item \revised{\textit{``ndev" (Number of Developers):}  It signifies the count of developers who have made changes to the modified files. Files altered by multiple developers tend to have a higher likelihood of containing defects, as suggested by~\citep {Matsumoto2010AnAO}. When ``ndev $>$ 107.00," it contributes to the prediction of "defective," consistent with the findings in Matsumoto et al.}

\item \revised{\textit{``nuc'' (Number of Unique Previous Changes):} It represents the number of unique changes within the modified files. Files with a history of numerous revisions are more prone to introducing defects because developers need to recall and track a multitude of past changes~\citep {DAmbros2010AnEC}. However, when ``nuc $>$ 553.00," its contribution to the ``clean" prediction is weaker, which slightly conflicts with the findings in DAmbros et al. It is possible that the ``nuc" value is not sufficiently large to strongly influence the ``defective" prediction.}

\item \revised{\textit{``ld''  (Deleted Code Lines):} It denotes the quantity of deleted code lines. Fewer deleted code lines are associated with a reduced likelihood of errors~\citep{Moser2008ACA, Nagappan2005UseOR}. When ``5.00$<$ld$<$19.00," it has an impact on the ``clean" prediction, in line with the observations in these studies.}

\item \revised{\textit{``lt'' (Code Lines Before Modification):} It stands for the number of code lines in the modified files before the changes were made. Larger files are more likely to contain defects~\citep{Koru2009AnII}. However, when ``lt $>$ 5520.75", it contributes weakly to the ``clean" prediction. This is possibly due to the relatively small ``lt'' values for these commits.}

\end{itemize}

\begin{figure}
    \centering
    \includegraphics[scale=0.21]{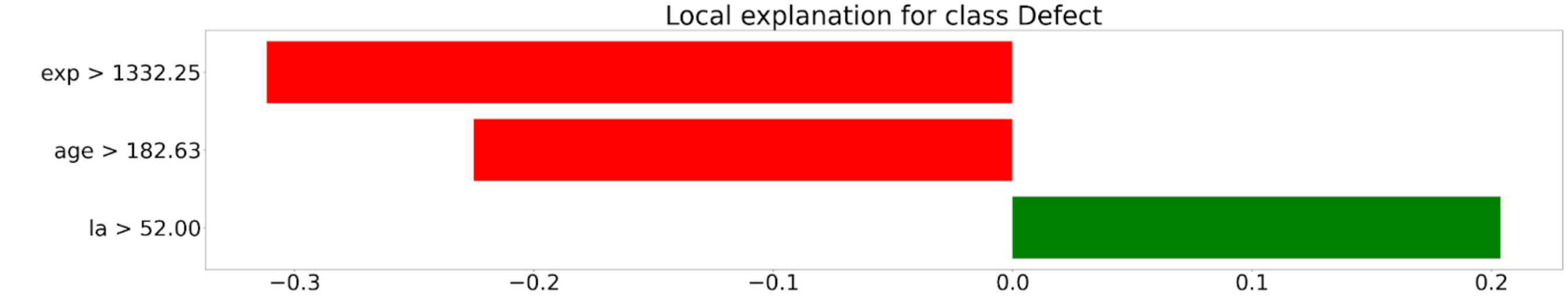}
    \vspace{-0.2cm}
  \caption{One example where Com predicts a ``clean'' commit as ``defective'' and Sim generates the correct prediction. The Chart is the explanations of Sim on its prediction by adopting the local explanation tool LIME~\citep{LIME}.}
  \label{rq4_qualtatitive_analysis_additional_example1}
\end{figure}

For a deep learning-based model like \emph{Com}, it is less likely to perceive those three top-ranked reasons solely from the code change tokens or commit log tokens. Thus, \emph{Com} could make mistakes without that prior knowledge.  
In summary, we observe that fusing expert knowledge with deep learning models could provide additional relevant knowledge to deep learning models and correct potential errors in their initial predictions.

\begin{figure}
    \centering
    \includegraphics[scale=0.25]{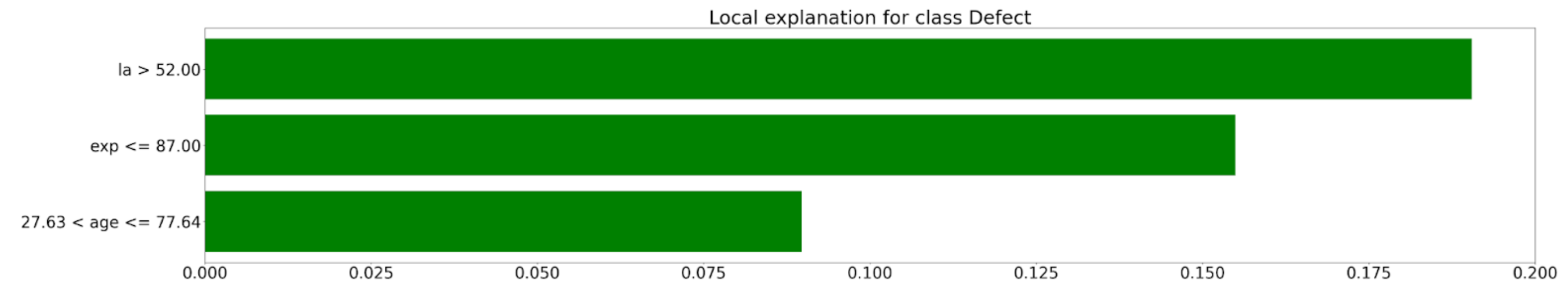}
    \vspace{-0.2cm}
  \caption{One example where Com predicts a ``defective'' commit as ``clean'' and Sim generates the correct prediction. The Chart is the explanations of Sim on its prediction by adopting the local explanation tool LIME~\citep{LIME}.}
  \label{rq4_qualtatitive_analysis_additional_example2}
\end{figure}

\revised{
To further demonstrate the benefits of combining Sim with Com, we present two additional examples. Figures~\ref{rq4_qualtatitive_analysis_additional_example1} and~\ref{rq4_qualtatitive_analysis_additional_example2} each refer to one of the two examples.}

\revised{As shown in Figure~\ref{rq4_qualtatitive_analysis_additional_example1}, 
in this example, the deep learning-based model Com initially predicts the commit as “defective”. However, in fact, this commit is a “clean” commit. Sim makes a correct prediction and SimCom++ also makes a correct prediction with the help of model fusion. 
After using LIME to explain the prediction of SIM, we found that ``exp'', ``age'', and ``la'' make substantial contributions to the prediction. Specifically, ``exp'' and ``age'' make contributions towards “Clean” (the correct prediction). ``la'' makes a contribution to “Defect” (the wrong prediction). Heuristics/expert knowledge behind those features are described below:}

\begin{itemize} [leftmargin=*]
    \item \revised{\textit{``exp'' (Developer Experience)}: It indicates how experienced a developer is. The more experienced developer (``exp $>$ 1332.25'') has a lower probability of introducing the defects.}

    \item \revised{\textit{``age'' (Age of Developers)}: It considers the module where the code changes happen is old and has not been changed for a long time (``age $>$ 182.63''), which may have fewer flaws according to (Graves et al., 2000).}

    \item \revised{\textit{``la'' (Added Code Lines)}: It denotes the quantity of added code lines. This commit does not add many new code lines (``la $>$ 52.00''). The more new code added, the higher the possibility of making mistakes, c.f.,~\citep{Moser2008ACA, Nagappan2005UseOR}.}
\end{itemize}

\revised{As shown in Figure~\ref{rq4_qualtatitive_analysis_additional_example2}, 
in this example, the deep learning-based model Com initially predicts the commit as “clean”. However, in fact, this commit is a “defective” commit. Sim makes a correct prediction and SimCom++ also makes a correct prediction with the help of model fusion.
After using LIME to explain the prediction of SIM, we found that “la”, “exp”, and “age” make substantial contributions to the prediction. Specifically, all of the three features make contributions towards “Clean” (the correct prediction). Heuristics/expert knowledge behind those features are described below:}

\begin{itemize} [leftmargin=*]

    \item \revised{\textit{``la'' (Added Code Lines)}: It denotes the quantity of added code lines. This commit does not add many new code lines (“la $>$ 52.00”). The more new code added, the higher the possibility of making mistakes, c.f.,~\citep{Moser2008ACA, Nagappan2005UseOR}}
    
    \item \revised{\textit{``exp'' (Developer Experience)}: It indicates how experienced a developer is. The less experienced developer (“exp $<$ 87.00”) has a lower probability of introducing the defects.}

    \item \revised{\textit{``age'' (Age of Developers)}: It considers the module where the code changes happen is relatively new and has been changed in a short time before (“age $<$ 77.64”), which may have more flaws according to ~\citep{Graves2000PredictingFI}}

\end{itemize}

\begin{center}
\begin{tcolorbox}[colback=gray!10,
                  colframe=black,
                  width=8.5cm,
                  arc=1mm, auto outer arc,
                  boxrule=0.5pt,
                 ]
\textbf{Answers to RQ4:} The model fusion strategy can reduce the false positives and retain most of the true positives of each component model. It can also correct the wrong predictions produced by its component models. The model fusion strategy works because it can merge both the expert knowledge with the commit contents to make a comprehensive assessment of a commit.
\end{tcolorbox}
\end{center}
\section{Discussion}

In this section, we present several discussions on the findings of this paper.

\subsection{On the impact of potential biases toward JIT defect prediction}

\vspace{0.1cm}
\textbf{Bias in JIT defect prediction.}
Wan et al.~\citep{Wan2020PerceptionsEA} surveyed practitioners to discuss the drawbacks of defect prediction tools. Some practitioners are unwilling to adopt defect prediction tools because they think the tools report nothing new.
It is well-known that commits that have large numbers of deleted or added lines are more likely defective~\citep{Nagappan2006MiningMT,Kamei2010RevisitingCB}.
Thus, it is necessary to explore the effectiveness of our approach in finding \textit{surprising} defects (i.e. defects hidden in the small or medium size commits).
Besides, many contribution guidelines~\footnote{\url{https://developer.blackbaud.com/skyux/contribute/contribution-process/guidelines}}~\footnote{ \url{https://wiki.lyrasis.org/display/DSPACE/Code+Contribution+Guidelines}}~\footnote{\url{https://cakebuild.net/community/contributing/contribution-guidelines}} of open-source software already advise developers to make small changes in a commit
because large code changes may take a very long time to review and test.
There will be more and more small commits in the future.

Inspired by the observations above,
we aim to investigate the effectiveness of SimCom++ on small or medium size commits by using datasets that \textit{exclude large commits}. 
\revised{However, defining a clear threshold for what constitutes a ``large'' commit is not straightforward. To address this issue without resorting to arbitrary thresholds, we adopted the following alternative approach: Firstly, we sorted the commits by size, arranging them from largest to smallest. Then we removed the top X\% of the largest commits, where X could be set to various percentages, including 0\%, 10\%, 20\%, 30\%, and 40\%. The reason we didn't set X to 100\% is that it would result in the removal of all commits, which wouldn't provide meaningful data for our analysis. For a breakdown of the number of small and large commits, when we delete the top X\% of large commits, the large commits account for X\% of all commits, while the small commits constitute (100-X)\% of the total commits in the dataset.}
To better understand how the models' performance varies when dropping different ratios of large commits, we conduct experiments by dropping the top 10\%, 20\%, 30\%, and 40\% of large commits.
Please note that we drop the large commits in the entire dataset (i.e. the training, validation, and testing data.)

\begin{figure*}[t]
    \centering
    \subfigure{
        \includegraphics[width=0.45\textwidth]{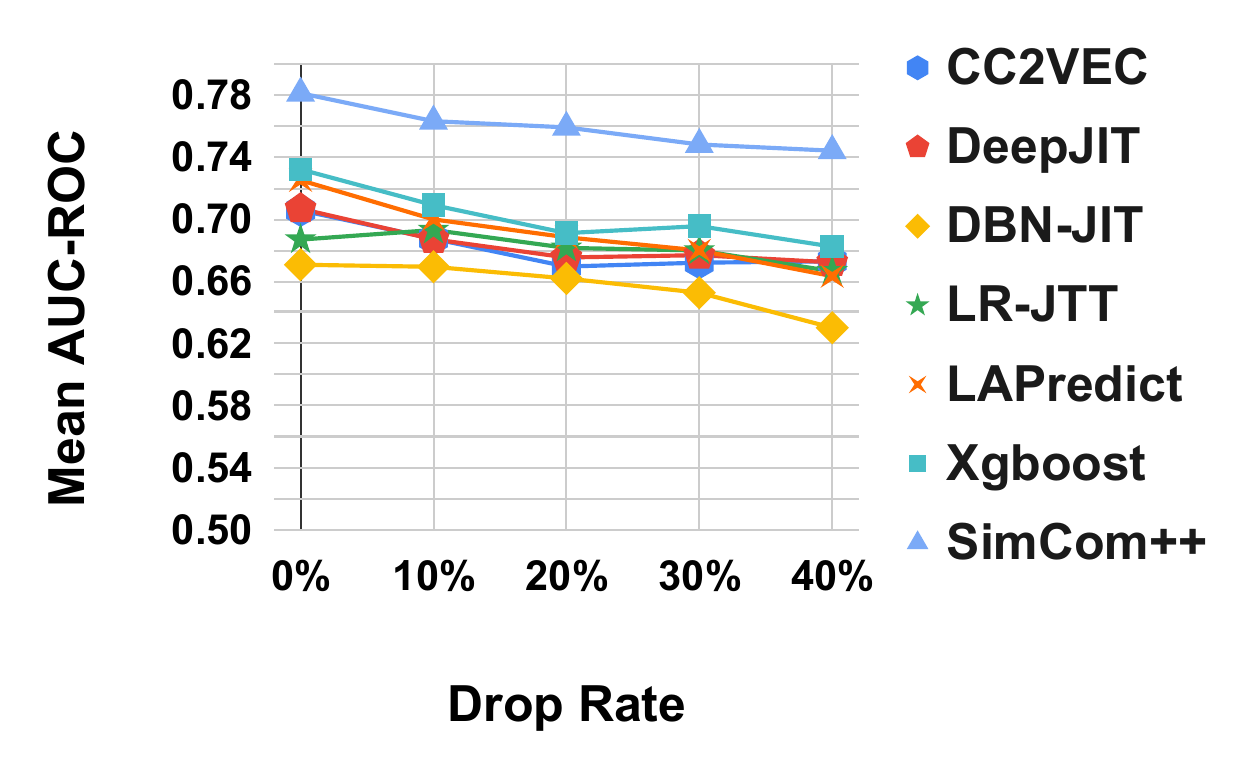}
        \label{T1}
    }
    \subfigure{
     \includegraphics[width=0.45\textwidth]{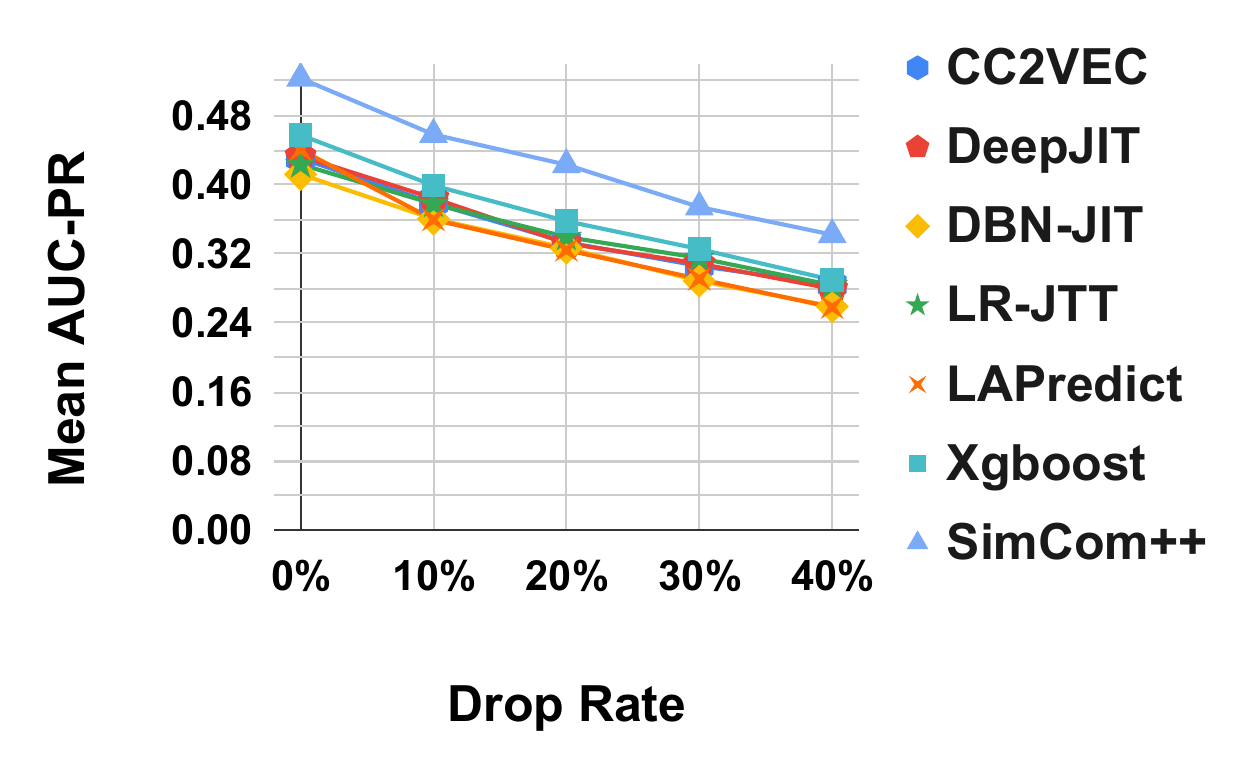}
        \label{P0}
    }
    \subfigure{
 \includegraphics[width=0.45\textwidth]{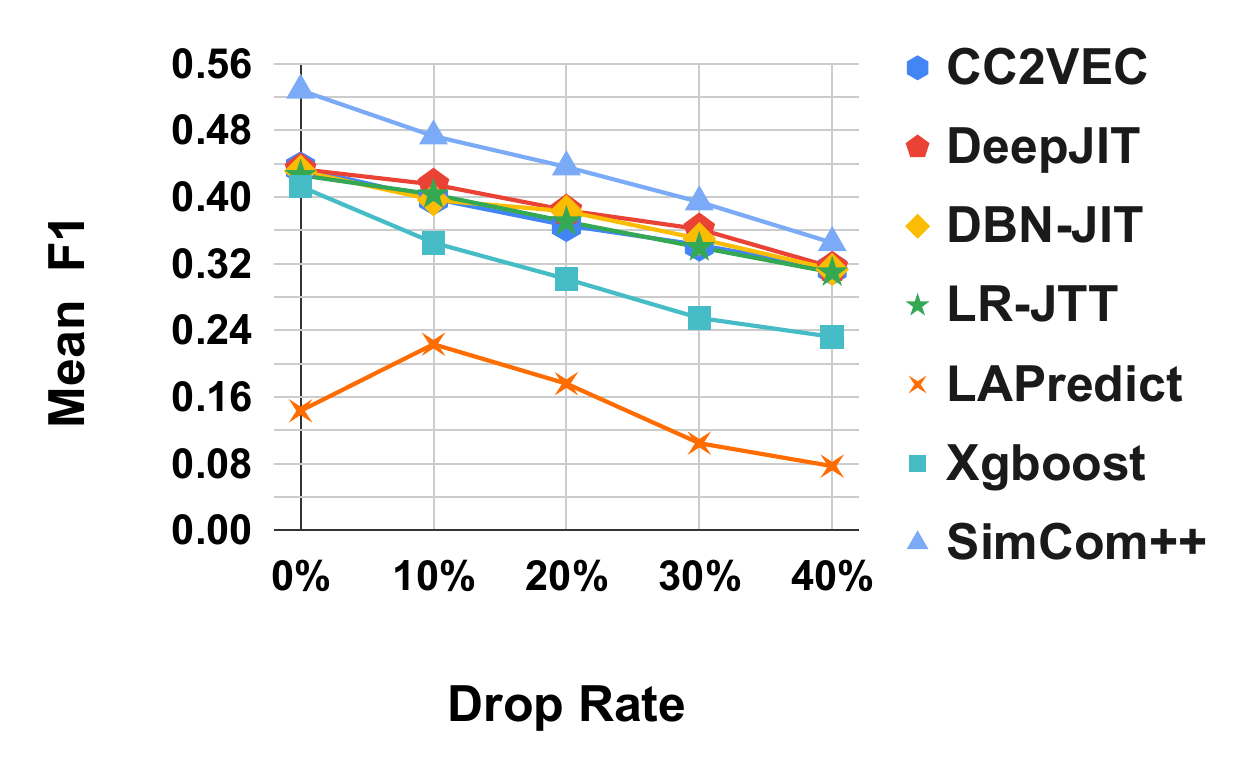}
        \label{F_}
    }
    \\
    \caption{The average performance of SimCom++ and baselines with different large commit drop rates.}
    \label{mean_performance}
\end{figure*}

\vspace{0.1cm}
\textbf{Results.}
Figure~\ref{mean_performance} shows the average performance 
of approaches in Setup II with respect to AUC-ROC, AUC-PR, and F1 scores, when dropping different ratios of largest commits. \revised{The average performance is determined by averaging the metrics from six projects: QT, OpenStack, JDT, Platform, Gerrit, and Go.} The y-axis in Figure~\ref{mean_performance} is the average performance and the x-axis is the ratio of the largest commits we drop. For instance, the drop rate equals 20\% indicating that we drop the top 20\% of large commits in datasets.

In general, comparing the performances in this special setting (excluding large commits) and the normal setting, all approaches perform worse in this special setting. One possible reason is that distinguishing buggy commits from clean commits in datasets that \textit{exclude large commits} is harder. 
Because over 40\% of those dropped large commits are defective commits on average, the number of buggy commits in this setting is smaller than that in the normal setting. Thus, it is easier to make false-positive predictions in this setting.
The experimental results show that SimCom++ consistently outperforms the best-performing baseline by 9.5\%, 16.1\%, and 13.5\% in terms of AUC-ROC, AUC-PR, and F1 scores on average across different drop rates.
This indicates that the superiority of SimCom++ over baselines does not change when the experimental setup changes.

In conclusion, when removing the large commits, SimCom++ still outperforms all baselines consistently and substantially on diverse evaluation metrics and projects.

\subsection{\revised{Stratified Cross Validation}}

\revised{In our previous experiments, we follow the previous work~\citep{Zeng2021DeepJD} to split the first 80\% of the datasets as the training sets and the regard the later 20\% of the datasets as the test sets. However, it brings limitations such as that we do not know whether the model performance is robust to other compositions of training/test sets. To address this limitation, we choose to evaluate our approach in a stratified cross-validation setting. Specifically, we conducted our experiments using a 5-fold stratified cross-validation setup.
Table~\ref{rq1_cross_validation} presents the model performance in the stratified cross-validation. We found SimCom++ consistently outperformed the best-performing baseline (i.e., XGBoost) by 6.8\%, 14.5\%, and 10.0\% in terms of AUC-ROC, AUC-PR, and F1, respectively.
In addition, SimCom++ also consistently outperformed its component models (Sim and Com). Specifically, SimCom++ outperforms Sim (Com) by 5.3\% (3.6\%), 10.5\% (6.3\%), and 7.0\% (3.6\%) in terms of AUC-ROC, AUC-PR, and F1, respectively.}

\revised{
We conducted the Wilcoxon Signed Rank at a 95\% significance level (i.e., p-value $<$ 0.05) and calculated Cliff’s delta on the paired data corresponding to SimCom++ and the best-performing baseline model (i..e, XGBoost). Our rationale is that if SimCom++ cannot significantly outperform any baseline, the best performer in each metric has the highest chance to be the baseline. 
The significance test has been conducted on the values of evaluation metrics. Specifically, when assessing F1 scores, we compute the F1 score for each sample within the test set and analyze the distribution of these scores. However, when dealing with AUC-ROC and AUC-PR, which measure the area under curves, it's crucial to have at least one sample from one class to derive a valid metric value. Consequently, it is not possible to calculate AUC-PR and AUC-ROC metrics for a single sample. To address this limitation, we randomly split the test set into 10 groups. Within each group, we calculate the AUC-ROC and AUC-PR metrics based on the available samples. This approach ensures that valid AUC-ROC and AUC-PR values are obtained. For Cliff’s delta, we consider a delta that is less than 0.147, between 0.147 and 0.33, between 0.33 and 0.474, and above 0.474 as Negligible (N), Small (S), Medium (M), and Large (L) effect size, respectively following previous literature~\citep{cliff2014ordinal}. We observed that p-values are less than 0.05 and Cliff’s delta are from 0.33 to 0.6, which correspond to the Medium or Large sizes. These results indicate that SimCom++ significantly and substantially outperforms the best-performing baseline model.
}

\begin{table}[]
 \centering
\caption{The performance of models in the fold-5 stratified cross-validation setting}
\resizebox{\columnwidth}{!}{%
\begin{tabular}{|l|l|r|r|r|r|r|r|r|}
\hline
\textbf{Tools}                    & \multicolumn{1}{c|}{\textbf{Metric}} & \multicolumn{1}{c|}{\textbf{QT}}       & \multicolumn{1}{c|}{\textbf{OP.}}      & \multicolumn{1}{c|}{\textbf{JDT}}      & \multicolumn{1}{c|}{\textbf{Plat.}}    & \multicolumn{1}{c|}{\textbf{Gerrit}}   & \multicolumn{1}{c|}{\textbf{GO}}       & \multicolumn{1}{c|}{\textbf{Avg.}}     \\ \hline
                                     & \cellcolor[HTML]{FFFFFF}ROC          & \cellcolor[HTML]{FFFFFF}0.650          & \cellcolor[HTML]{FFFFFF}0.708          & \cellcolor[HTML]{FFFFFF}0.656          & \cellcolor[HTML]{FFFFFF}0.682          & \cellcolor[HTML]{FFFFFF}0.766          & \cellcolor[HTML]{FFFFFF}0.634          & \cellcolor[HTML]{FFFFFF} 0.683          \\ \cline{2-9} 
                                     & \cellcolor[HTML]{EFEFEF}PR           & \cellcolor[HTML]{EFEFEF}0.274          & \cellcolor[HTML]{EFEFEF}0.456          & \cellcolor[HTML]{EFEFEF}0.630          & 
                                     \cellcolor[HTML]{EFEFEF}0.532          & \cellcolor[HTML]{EFEFEF}0.310          & \cellcolor[HTML]{EFEFEF}0.564          & \cellcolor[HTML]{EFEFEF}0.461          \\ \cline{2-9} 
\multirow{-3}{*}{\textbf{LR-JIT}}    & \cellcolor[HTML]{CBCEFB}F1           & \cellcolor[HTML]{CBCEFB}0.310          & \cellcolor[HTML]{CBCEFB}0.496          & \cellcolor[HTML]{CBCEFB}0.624          & \cellcolor[HTML]{CBCEFB}0.558          & \cellcolor[HTML]{CBCEFB}0.352          & \cellcolor[HTML]{CBCEFB}0.578          & \cellcolor[HTML]{CBCEFB}0.486          \\ \hline
& ROC                                  & 0.690                                  & 0.718                                  & 0.620                                  & 0.724                                  & 0.776                                  & 0.680                                  & 0.701                                  \\ \cline{2-9} 
                                     & \cellcolor[HTML]{EFEFEF}PR           & \cellcolor[HTML]{EFEFEF}0.284          & \cellcolor[HTML]{EFEFEF}0.472          & \cellcolor[HTML]{EFEFEF}0.598          & \cellcolor[HTML]{EFEFEF}0.578          & \cellcolor[HTML]{EFEFEF}0.336          & \cellcolor[HTML]{EFEFEF}0.612          & \cellcolor[HTML]{EFEFEF}0.480          \\ \cline{2-9} 
\multirow{-3}{*}{\textbf{DBN-JIT}}   & \cellcolor[HTML]{CBCEFB}F1           & \cellcolor[HTML]{CBCEFB}0.362         & \cellcolor[HTML]{CBCEFB}0.502          & \cellcolor[HTML]{CBCEFB}0.540          & \cellcolor[HTML]{CBCEFB}0.592          & \cellcolor[HTML]{CBCEFB}0.402          & \cellcolor[HTML]{CBCEFB}0.578          & \cellcolor[HTML]{CBCEFB}0.496          \\ \hline
& ROC                                  & 0.738                                  & 0.744                                  & 0.674                                  & 0.722                                  & 0.766                                  & 0.708                                 &0.725                                  \\ \cline{2-9} 
                                     & \cellcolor[HTML]{EFEFEF}PR           & \cellcolor[HTML]{EFEFEF}0.324          & \cellcolor[HTML]{EFEFEF}0.478          & \cellcolor[HTML]{EFEFEF}0.630          & \cellcolor[HTML]{EFEFEF}0.594          & \cellcolor[HTML]{EFEFEF}0.286          & \cellcolor[HTML]{EFEFEF}0.512          & \cellcolor[HTML]{EFEFEF}0.487          \\ \cline{2-9} 
\multirow{-3}{*}{\textbf{LApredict}} & \cellcolor[HTML]{CBCEFB}F1           & \cellcolor[HTML]{CBCEFB}0.006          & \cellcolor[HTML]{CBCEFB}0.058          & \cellcolor[HTML]{CBCEFB}0.416          & \cellcolor[HTML]{CBCEFB}0.158          & \cellcolor[HTML]{CBCEFB}0.084          & \cellcolor[HTML]{CBCEFB}0.028          & \cellcolor[HTML]{CBCEFB}0.125          \\ \hline
& ROC                                  & 0.710                                  & 0.710                                  & 0.604                                  & 0.754                                  & 0.794                                  & 0.690                                  & 0.710                                  \\ \cline{2-9} 
                                     & \cellcolor[HTML]{EFEFEF}PR           & \cellcolor[HTML]{EFEFEF}0.278          & \cellcolor[HTML]{EFEFEF}0.433          & \cellcolor[HTML]{EFEFEF}0.558          & \cellcolor[HTML]{EFEFEF}0.590          & \cellcolor[HTML]{EFEFEF}0.410          & \cellcolor[HTML]{EFEFEF}0.598          & \cellcolor[HTML]{EFEFEF}0.479          \\ \cline{2-9} 
\multirow{-3}{*}{\textbf{DeepJIT}}   & \cellcolor[HTML]{CBCEFB}F1           & \cellcolor[HTML]{CBCEFB}0.350          & \cellcolor[HTML]{CBCEFB}0.493          & \cellcolor[HTML]{CBCEFB}0.626          & \cellcolor[HTML]{CBCEFB}0.640          & \cellcolor[HTML]{CBCEFB}0.366          & \cellcolor[HTML]{CBCEFB}0.642          & \cellcolor[HTML]{CBCEFB}0.519          \\ \hline
& ROC                                  & 0.710                                  & 0.718                                  & 0.610                                  & 0.750                                  & 0.794                                  & 0.692                                  & 0.712                                  \\ \cline{2-9} 
                                     & \cellcolor[HTML]{EFEFEF}PR           & \cellcolor[HTML]{EFEFEF}0.284          & \cellcolor[HTML]{EFEFEF}0.433          & \cellcolor[HTML]{EFEFEF}0.562          & \cellcolor[HTML]{EFEFEF}0.588          & \cellcolor[HTML]{EFEFEF}0.408          & \cellcolor[HTML]{EFEFEF}0.598          & \cellcolor[HTML]{EFEFEF}0.479          \\ \cline{2-9} 
\multirow{-3}{*}{\textbf{CC2Vec}}    & \cellcolor[HTML]{CBCEFB}F1           & \cellcolor[HTML]{CBCEFB}0.310          & \cellcolor[HTML]{CBCEFB}0.478          & \cellcolor[HTML]{CBCEFB}0.578          & \cellcolor[HTML]{CBCEFB}0.638          & \cellcolor[HTML]{CBCEFB}0.302          & \cellcolor[HTML]{CBCEFB}0.640          & \cellcolor[HTML]{CBCEFB}0.491          \\ \hline
& ROC                                  & 0.715                                  & 0.735                                  & 0.610                                  & 0.766                                  & 0.818                                 & 0.720                                  & 0.727                                 \\ \cline{2-9} 
& \cellcolor[HTML]{EFEFEF}PR           & \cellcolor[HTML]{EFEFEF}0.315         & \cellcolor[HTML]{EFEFEF}0.483          & \cellcolor[HTML]{EFEFEF}0.564          & \cellcolor[HTML]{EFEFEF}0.616          & \cellcolor[HTML]{EFEFEF}0.403          & \cellcolor[HTML]{EFEFEF}0.636          & \cellcolor[HTML]{EFEFEF}0.503          \\ \cline{2-9} 
\multirow{-3}{*}{\textbf{CodeBERT}}  & \cellcolor[HTML]{CBCEFB}F1           & \cellcolor[HTML]{CBCEFB}0.340          & \cellcolor[HTML]{CBCEFB}0.490          & \cellcolor[HTML]{CBCEFB}0.570          & \cellcolor[HTML]{CBCEFB}0.598          & \cellcolor[HTML]{CBCEFB}0.458          & \cellcolor[HTML]{CBCEFB}0.616          & \cellcolor[HTML]{CBCEFB}0.512          \\ \hline
& ROC                                  & 0.748                                  & 0.758                                  & 0.668                                  & 0.780                                  & 0.770                                  & 0.744                                  & 0.745                                  \\ \cline{2-9} 
                                     & \cellcolor[HTML]{EFEFEF}PR           & \cellcolor[HTML]{EFEFEF}0.356          & \cellcolor[HTML]{EFEFEF}0.516          & \cellcolor[HTML]{EFEFEF}0.642          & \cellcolor[HTML]{EFEFEF}0.628          & \cellcolor[HTML]{EFEFEF}0.366          & \cellcolor[HTML]{EFEFEF}0.682          & \cellcolor[HTML]{EFEFEF}0.532          \\ \cline{2-9} 
\multirow{-3}{*}{\textbf{XGBoost}}   & \cellcolor[HTML]{CBCEFB}F1           & \cellcolor[HTML]{CBCEFB}0.394          & \cellcolor[HTML]{CBCEFB}0.536          & \cellcolor[HTML]{CBCEFB}0.594          & \cellcolor[HTML]{CBCEFB}0.632          & \cellcolor[HTML]{CBCEFB}0.346          & \cellcolor[HTML]{CBCEFB}0.646          & \cellcolor[HTML]{CBCEFB} 0.525        \\ \hline
& ROC                                  & 0.746                                  & 0.764                                  & 0.588                                  & 0.788                                  & 0.808                                  & 0.752                                  & 0.756                                  \\ \cline{2-9} 
                                     & \cellcolor[HTML]{EFEFEF}PR           & \cellcolor[HTML]{EFEFEF}0.344          & \cellcolor[HTML]{EFEFEF}0.526          & \cellcolor[HTML]{EFEFEF}0.662          & \cellcolor[HTML]{EFEFEF}0.630          & \cellcolor[HTML]{EFEFEF}0.460          & \cellcolor[HTML]{EFEFEF}0.685          & \cellcolor[HTML]{EFEFEF}0.551          \\ \cline{2-9} 
\multirow{-3}{*}{\textbf{Sim}}       & \cellcolor[HTML]{CBCEFB}F1           & \cellcolor[HTML]{CBCEFB}0.398          & \cellcolor[HTML]{CBCEFB}0.542          & \cellcolor[HTML]{CBCEFB}0.618          & \cellcolor[HTML]{CBCEFB}0.648          & \cellcolor[HTML]{CBCEFB}0.388          & \cellcolor[HTML]{CBCEFB}0.648          & \cellcolor[HTML]{CBCEFB}0.540          \\ \hline
 & ROC                                  & 0.750                                  & 0.772                                  & 0.688                                  & 0.772                                  & 0.848                                  & 0.776                                  & 0.768                                  \\ \cline{2-9} 
                                     & \cellcolor[HTML]{EFEFEF}PR           & \cellcolor[HTML]{EFEFEF}0.366          & \cellcolor[HTML]{EFEFEF}0.550          & \cellcolor[HTML]{EFEFEF}0.668          & \cellcolor[HTML]{EFEFEF}0.616          & \cellcolor[HTML]{EFEFEF}0.535          & \cellcolor[HTML]{EFEFEF}0.704          & \cellcolor[HTML]{EFEFEF}0.573          \\ \cline{2-9} 
\multirow{-3}{*}{\textbf{Com}}       & \cellcolor[HTML]{CBCEFB}F1           & \cellcolor[HTML]{CBCEFB}0.394          & \cellcolor[HTML]{CBCEFB}0.536          & \cellcolor[HTML]{CBCEFB}\textbf{0.640}          & \cellcolor[HTML]{CBCEFB}0.646          & \cellcolor[HTML]{CBCEFB}0.458          & \cellcolor[HTML]{CBCEFB}0.674          & \cellcolor[HTML]{CBCEFB}0.558          \\ \hline
& ROC                                  & \textbf{0.774}                         & \textbf{0.803}                         & \textbf{0.710}                         & \textbf{0.818}                         & \textbf{0.878}                         & \textbf{0.795}                         & \textbf{0.796}                         \\ \cline{2-9} 
                                     & \cellcolor[HTML]{EFEFEF}PR           & \cellcolor[HTML]{EFEFEF}\textbf{0.402} & \cellcolor[HTML]{EFEFEF}\textbf{0.598} & \cellcolor[HTML]{EFEFEF}\textbf{0.694} & \cellcolor[HTML]{EFEFEF}\textbf{0.680} & \cellcolor[HTML]{EFEFEF}\textbf{0.545} & \cellcolor[HTML]{EFEFEF}\textbf{0.738} & \cellcolor[HTML]{EFEFEF}\textbf{0.609} \\ \cline{2-9} 
 \multirow{-3}{*}{\textbf{SimCom++}}  & \cellcolor[HTML]{CBCEFB}F1           & \cellcolor[HTML]{CBCEFB}\textbf{0.422} & \cellcolor[HTML]{CBCEFB}\textbf{0.580} & \cellcolor[HTML]{CBCEFB}0.638 & \cellcolor[HTML]{CBCEFB}\textbf{0.678} & \cellcolor[HTML]{CBCEFB}\textbf{0.465} & \cellcolor[HTML]{CBCEFB}\textbf{0.693} & \cellcolor[HTML]{CBCEFB}\textbf{0.578} \\ \hline
\end{tabular}
\label{rq1_cross_validation}
}
\end{table}

\subsection{Implication}

Simple models and complex models can be partners rather than adversaries.
In prior work on JIT defect prediction, most works focused on either how to make better use of hand-made features by traditional ML classifiers, or how to effectively extract features from the commit content by DL techniques. As there is a big gap between commit-level hand-made features (a list of numbers) and commit contents (commit message tokens and code tokens), there is little work to leverage these two distinct input data.
Our results indicate that combining both the simple model (dealing with expert knowledge) and the complex model (dealing with commit contents) is promising in JIT defect prediction. 
Our work suggests that future work should explore further this idea on other software engineering tasks that have both meaningful hand-made features and effective DL models.

\subsection{Threats to Validity}
\vspace{0.2cm}\noindent{\bf Threats to Internal Validity.} 
For all baselines considered in this study, we directly use the source code in the replication packages of the studied techniques and use the default hyperparameters.
To reduce the threat, we carefully reviewed the experimental scripts to ensure their correctness.
We also release our replication package for others to check.

\vspace{0.1cm}\noindent{\bf Threats to External Validity.}
Threats to external validity are concerned with the generalizability of our findings.
To reduce the threat, we use a diverse and large-scale JIT defect prediction dataset of real-world projects in different programming languages collected by Zeng et al.~\citep{Zeng2021DeepJD}.    
As this dataset contains a large number of labeled commits from six projects,
our findings on this larger-scale dataset have
the generalizability to other projects to some extend. 
\revised{Another threat to external validity lies in the generalizability of our approach in different data splitting strategies. In line with previous work~\citep{Zeng2021DeepJD}, we have employed the 80\%-20\% training/test set split for evaluation. However, our model may not perform well if the data splitting changes. To mitigate this threat, we did experiments using a stratified cross-validation setup with 5 folds. Table~\ref{rq1_cross_validation} showed that our proposed model SimCom++ still significantly outperformed the baselines by a large margin when the compositions of training/test set changes. Thus, our model shows its robustness to data splitting changes to some extent.
}

\vspace{0.1cm}\noindent{\bf Threats to Construct Validity.}  
One of the main threats to construct validity is the evaluation metrics we choose.
To mitigate this threat, following the DeepJIT and LApredict studies, 
we adopt the widely-used AUC-ROC score to evaluate the performance.
As ROC curves can lead to an overly optimistic view of a model’s performance on skewed datasets~\citep{Davis2006TheRB}, we also adopt an alternative metric for AUC-ROC (i.e. AUC-PR) that is more suitable for the skewed datasets.
Besides, to evaluate the ability of the approaches in making correct predictions, we also consider the F1-score, which is computed based on the number of true/false positive/negatives.

\section{Related Work}

Many approaches have been proposed for JIT defect prediction. 
Mockus et al.~\citep{Mockus2000PredictingRO} built a model by using the historic information in commits and use the model to predict the risk of new commits. 
Kamei et al.~\citep{Kamei2013ALE} built an effort-aware prediction model using Logistic Regression with 14 change-level, hand-crafted metrics (features). 
Yang et al.~\citep{Yang2015DeepLF} applied Deep Belief Network (DBN) to extract higher-level information from the change-level metrics.
Later, Yang et al.~\citep{Yang2017TLELAT} proposed a two-layer ensemble learning model that combines decision tree and ensemble learning. 
This two-layer model achieved better performance for JIT defect prediction. 
Young et al.~\citep{Young2018ARS} proposed a new ensemble approach by using arbitrary classifiers and optimizing the weights of the classifiers. 
Liu et al.~\citep{Liu2017CodeCA} proposed an unsupervised learning approach based on code churn in effort-aware settings.
Cabral et al.~\citep{Cabral2019ClassIE} proposed a new sampling method to address the issues of verification latency and class imbalance evolution in the online JIT defect prediction setting. 
Then, Yan et al.~\citep{Yan2020JustInTimeDI} proposed a two-phase framework that can handle the identification and localization tasks at the same time. 
Recently, Hoang et al.~\citep{Hoang2019DeepJITAE} proposed DeepJIT, which uses deep learning techniques to extract the features automatically from commit logs and code changes.
To enhance it, Hoang et al.~\citep{Hoang2020CC2VecDR} proposed a new approach that learns the representation of code changes through the supervision of commit logs. Recently, Zeng et al.~\citep{Zeng2021DeepJD} conducted a study on the deep learning-based JIT defect prediction models (i.e. DeepJIT and CC2Vec) on an extended dataset and found that deep learning-based approaches cannot outperform a simple model namely LApredict.


Most of the existing work focused on either how to make better use of hand-made features by traditional ML classifiers, or how to extract better features from the commits by DL techniques. 
Different from the prior work, our model combines the simple model with the complex model by early and late fusion techniques and outperforms the baselines significantly.
Our results indicate that it is promising to combine the simple model (expert knowledge) and the complex model (commit contents with DL methods) to achieve better performance.

\revised{Another closely related work is DBN-JIT~\citep{Yang2015DeepLF}, which uses a deep learning model with hand-crafted features. Our proposed approach SimCom++ is different from DBN-JIT in the following aspects: 
Firstly, the models have different feature sources. In DBN-JIT, feature extraction primarily involves the use of deep learning techniques to pick features from an existing set of hand-crafted features. In contrast, SimCom++ utilizes deep learning techniques (specifically, the Com model of SimCom++) to automatically extract new features from the contents of commits. Secondly, the models have different structures. DBN-JIT is a single model for JIT defect prediction. In contrast, SimCom++ represents the amalgamation of multiple models, strategically combined to achieve enhanced effectiveness.
}

\section{Conclusion and future work}

In this work, we propose a framework that effectively combines one simple model and one complex model via both early and late fusion.
The experimental results show that our approach can significantly outperform the baselines and the state-of-the-art by 5.7\%, 12.5\%, and 17.9\% in terms of AUC-ROC, AUC-PR, and F1-score on average.
Besides, we found that our framework (combining simple and complex models) performs consistently better (4.4\%--10.1\% on average) than the simple model only or the complex model only in all different evaluation metrics and software projects.
To further evaluate the effectiveness of our approach in different settings, we also explore the performance of JIT defect predictors in another setting that excludes large commits (i.e. top 10\%--40\% large commits): our approach can outperform the best-performing baselines by a large improvement (9.5\%, 16.1\%, and 13.5\% in terms of AUC-ROC, AUC-PR, and F1-score) in this setting, indicating the effectiveness of our approach in different settings.
In the future, we are interested in investigating the effectiveness of combining simple models and complex models in other SE tasks.



\section{Data Availability}
We share the replication package\footnote{\revised{\url{https://anonymous.4open.science/r/SimCom--F5F1/}}} that contains the code and datasets used in this study for further evaluation and extension of our study.


%
\section*{Declarations}

\textbf{Conflict of Interests} The authors have no conflicts of interest to declare that are relevant to this paper.

\bibliographystyle{spbasic}      
\bibliography{bib.bib}   


\end{document}